\documentclass[structabstract]{aa}  
\usepackage{graphicx}
\usepackage{savesym}
\usepackage{amsmath}
\usepackage{amssymb}
\usepackage{txfonts}
\usepackage{time}
\usepackage[usenames]{color}

\usepackage{natbib}
\bibpunct{(}{)}{;}{a}{}{,}

\begin{document}
\newcommand{\fix}[1]{{#1}}
\newcommand{\lf}[1]{{#1}}
\newcommand{\lff}[1]{{#1}}
\bibliographystyle{aa}
\frenchspacing
\title{Spatially coupled inversion of spectro-polarimetric image data I: Method and first results}
\titlerunning{Inversion of \lf{spectro-polarimetric} image data}
 
\author{M.\ van Noort}

\date{\today}
                  
\authorrunning{M. van Noort}

\offprints{M. v. N.: \email{vannoort@mps.mpg.de}}

\institute{Max-Planck Institute for Solar System Research,
  Max-Planck Stra\ss e 1,
  D-37412 Katlenburg-Lindau, Germany}

\abstract
{When inverting solar spectra, image degradation effects that are present in the data are usually approximated or not considered.}
{\lf{We} develop a data reduction method that takes these issues into account and minimizes the resulting errors.}
{By accounting for the diffraction PSF of the telescope during the inversions, we can produce a self-consistent solution that best fits the observed data, \lf{while simultaneously}  requiring fewer free parameters than conventional approaches.}
{Simulations \lf{using} realistic MHD data indicate that the method is stable for all resolutions, including those with pixel scales well beyond those that can be resolved with a 0.5m telescope, such as the Hinode SOT. Application of the presented method to reduce full Stokes data from the Hinode spectro-polarimeter results in dramatically increased image contrast and an increase in the resolution of the data to the diffraction limit of the telescope in almost all Stokes and fit parameters. The resulting data allow for \lf{detecting} and \lf{interpreting} solar features that have so far only been observed with 1m class ground-based telescopes.}
{A new inversion method was developed that allows for accurate fitting of solar spectro-polarimetric imaging data over a large field of view, while simultaneously improving the noise statistics and spatial resolution of the results \lf{significantly}.}


\maketitle

\section{Introduction}
\lf{In solarphysical research, as in the vast majority of astrophysical research,} information about the object under investigation is received in the form of electromagnetic radiation. \lf{Since} this radiation was emitted by, passed through or reflected off the object of interest, \lf{interpreting} the information contained in the radiation typically involves \lf{computing} the observed data from a model of the object using basic physical principles.

\lf{Interpreting} the data typically starts by making an educated guess about the observed object\lff{'}s atmosphere \lff{(the model)}, followed by an evaluation of the compatibility of the observable\lff{,} calculated using this model\lff{,} with the observed data. Additional improvements may then be made, depending on the flexibility of the model. Optimizing the compatibility of the synthesized results with the data is usually conveniently formulated in terms of the minimization of the square of the differences. A number of codes exist that automate this optimization process, either by a standard linearized optimization (SIR \citep{1992ApJ...398..375R}, SPINOR \citep{2000PhDTh,2000A&A...358.1109F}, NICOLE \fix{\citep{1998ApJ...507..470S}}, HAZEL \citep{2008ApJ...683..542A}, and many more)\lff{;} by statistical methods (HELIX \citep{2004A&A...414.1109L,2009ASPC..415..327L}, MERLIN, etc.)\lf{;} by means of Bayesian methods \lff{\citep[Bayes-ME][]{2007A&A...476..959A};} or by using artificial neural networks \fix{\citep{2003NN.....16..355S,2008A&A...488..781C}}, with greatly varying levels of detail in their treatment of the physics \lf{and} often with the aim of treating very specific types of data.

Usually the simplest atmospheric model that can reproduce the observations is used for analysis of the solar lower atmosphere. The model is characterized by a number of nodes, at which the \fix{physical parameters that are relevant for the formation of the spectrum are defined. These  quantities are then interpolated to a finer grid to allow for \lf{the calculation of} an accurate solution \lf{to} the radiative transfer equation}. Automatic inversion codes, \lf{which optimize} this kind of model\lff{,} have been in use for two decades for \lf{interpreting} solar data.

\lf{Spurred by} the increase in computing power in recent times, much effort has gone into increasing the level of detail and self-consistency in the treatment of the basic physics, non-LTE (NICOLE) and scattering polarization effects (HELIX, HAZEL, etc.). \lf{Without these,} many observed spectral features can be shown to be poorly reproduced, even in cases where the fitted atmosphere is realistic.

The observed data used in this process, however, quantify the properties of radiation that has passed through a telescope, an instrument\lf{,} and a detector, all of which alter the data depending on the size, optical quality and other instrumental properties. For ground-based observations, in addition, the radiation passes through the earths atmosphere, which redistributes it \lf{by} refracting and scattering.

A large body of work exists on removing the degradation of the spatial resolution due to \lf{an} increase in the \lf{extent} of the point spread function from instrumental and atmospheric effects. In particular the optimization of image information from observed data degraded by atmospheric turbulence ((MO)MFBD \citep{2005SoPh..228..191V}, phase diversity methods \lf{\citep[][and references therein]{paxman92joint,1994A&AS..107..243L},} and speckle reconstruction methods \lf{\citep[][and many more]{1970A&A.....6...85L,1974ApJ...193L..45K,1983ApOpt..22.4028L,1992A&A...257L...4D,1993A&A...268..374V}}) has recently found widespread application in spectral imaging data produced by an array of narrow band filter instruments; e.g. SOUP \citep{title81SOUP}, CRISP \citep{2006A&A...447.1111S}, GFPI \citep{2006A&A...451.1151P} \lf{and} IBIS \citep{2006SoPh..236..415C}, to name just a few. These methods only work well for variable degradation on timescales [much] shorter than the time resolution of the observation.

In the spectral dimension\lf{,} on the other hand, degradation by instrumental and atmospheric effects has a rather different effect on the data. Where the spatial degradation merely decreases the resolution in the spatial dimensions, the blending of the spectra results in observed spectra containing a mix of information from a collection of different atmospheric components. Much work \lf{has been} done on the analysis and interpretation of such ``multi-component'' spectra \lf{\citep[][and many more]{1968AULII.....2..2S,1991sopo.work..307S,1996SoPh..164..277B}}, by modeling them as arising from two or more separate atmospheric ``components'', \lf{which} remain unresolved in the data. 

\cite{2007PASJ...59S.837O} model the effect of telescope diffraction on the spectra as a local ``stray-light'' contribution, where the light from the surroundings is taking the place of the contribution of a second atmospheric component. The apparent multi-component spectra are not interpreted in terms of unresolved fine structure, but rather in terms of an additive contribution produced by telescope diffraction.

In this paper we develop this idea \lf{further} and develop a method that rigorously deals with the effect of image degradation by instrumental and telescope effects, by integrating these effects in a 2-D adaptation of the SPINOR inversion code.

\section{Inversion of spatially degraded data}
Any process that infers the most probable atmospheric structure that gave rise to the observed spectral data may generally be referred to as an inversion process. One could even argue that the inversion is \lf{only defined} by the way the probability of a particular atmospheric structure is defined, since the method used to locate the point of maximum likelihood is normally quite independent \lf{of} the formulation of the probability function, but this would not do justice to the information that must be obtained from the atmospheric structure in order to optimize it. 

Considerable differences exist between the various inversion codes in the way they describe the atmosphere, the description of the radiative transfer problem\lf{,} and in the way they try to maximize the likelihood function. However, what they all have in common is that ultimately they all compute synthetic profiles from estimated atmospheres and compare these to the observed profile, in order to infer the most likely atmosphere that produced the observed spectrum.

A physically accurate description is included in most of these code\lf{s} in ample detail, \lf{since} it is obvious that this is needed to have any chance \lf{of interpreting} an observed spectrum in terms of physical quantities. Although the most common assumption made in inversion codes is that of local thermodynamic equilibrium (LTE), there are codes \lf{that compute} the line profiles in full NLTE (e.g. NICOLE, HAZEL), \lf{codes that} include detailed scattering physics (HELIX, HAZEL)\lff{,} and \lf{codes that include detailed} molecular physics (SPINOR). In addition, there is a great variation in the amount of detail in which the atmosphere is parameterized, from slabs with constant properties to full \lf{3-D} structures with a full height stratification in all physical quantities \lff{\citep[see for instance][for an overview]{2012ApJ...748...83A}}.

Instrumental effects, on the other hand, with the exception of spectral smearing, are not generally considered in much detail, if at all, by many codes, \lf{because} this is considered by most code authors to be taken care of by appropriate data reduction routines. Although careful calibration of the data generally receives a \lf{lot} of attention, there are always significant limits to the extent to which it is possible to remove the instrumental effects from the data. 

In particular, the removal from the data of any degradation of the data that reduces the information content (such as convolution with an extended convolution kernel) will result in amplification of any additional modification that was made to the data {\em after} that degradation took place, regardless of whether this modification is in the form of detector or photon noise, detector \lf{nonlinearities}, or inhomogeneities in the response of any of the elements taking part in the acquisition of the data (such as filters or modulators). For this reason, it is advantageous to apply the instrumental effects to the synthetic data as a part of the relevant physics, rather than attempting to remove them from the data itself. 
The most common instrumental degradation treated in this way so far is the degradation in the spectral dimension. Since the effect of this degradation on the spectra is frequently very severe, convolution of the synthesized spectra with the instrumental transmission profile before comparing to the observed data is a standard ingredient in almost any inversion code. 

A rudimentary treatment of spatial degradation effects can be found in the form of a micro- and/or macro-turbulent broadening of the spectral line profiles in almost all inversion codes\lf{;} however, a thorough treatment of degradations in the spatial domain has thus far not been employed. This is probably not due to the relative importance of spatial degradation as compared to spectral degradation, but more likely due to the limitations imposed by the available datasets. Recent advances in instrumentation (Hinode \lf{SP}, IBIS, CRISP, etc), have produced datasets sampling both the spatial and the spectral dimensions \lf{with high enough resolution} to open up the possibility \lf{of} improved treatment. 

\cite{2007PASJ...59S.837O} used the observed data from neighboring pixels to assemble a ``local stray-light'' spectrum, \lf{which} contributes to the observed spectrum in a particular pixel of Hinode SP slit\lf{-}scan\lf{ned} data. As acknowledged by the authors, this method is far from exact, but is simple to understand \lf{and} easy to implement\lf{,} and \lf{it} fixes most of the fitting problems they encountered. 

However, since the estimate of the ``stray-light'' contribution can only be calculated from the observed data, this method in effect approximates a deconvolution \lf{by subtracting} a running average. Apart from being formally incorrect, this procedure has the notable disadvantage that it removes an estimate of the signal not originating in a particular pixel (local stray-light) from the observations, thereby substantially decreasing the \lf{signal-to-noise} ratio of the data. In addition, the process of using the data to calculate a fit parameter (the stray-light fraction) makes the solution vulnerable to systematic errors \citep{2011ApJ...731..125A}.

In the following, we develop an integrated approach to the problem of \lf{inverting} spatially and spectrally degraded data. To see how to do this, we must first analyze the problem in some more detail.

\subsection{Image response}
\label{sec:imaging}

The collection of solar spectral imaging data is no different from any other imaging problem, in that an image of the solar surface is formed by a telescope through an aperture in a focal plane, where it is sampled by an instrument that can detect both the spatial and spectral variability of the incident radiation. As a consequence of the finite aperture of any imaging device, the radiation originating in a point on the object is redistributed across a spatially extended region \lf{on} the detector.

The implication of this for two neighboring points \lf{of} the object is that their respective spectra \lf{have been averaged} before they \lf{are} detected in the focal plane. In the probable case that the spectra differ, the result is a mixed spectrum that may not even resemble the original spectra.

To describe the effect of the spatial smearing with a PSF $\phi(x,y,\lambda)$, we consider the data observed if only a single point source \lf{is} present in $(x_0,y_0)$:
$$
d(x,y,\lambda)= i(x_0,y_0,\lambda) \phi(x-x_0,y-y_0,\lambda).
$$
For a spectrum that is completely described by a set of parameters $\alpha_i$ we may then write for the response of the observed data to a change of the parameter $\alpha_i$ in location $(x,y)$:
\begin{equation}
\frac{\partial d(x,y,\lambda)}{\partial\alpha_i}=\phi(x-x_0,y-y_0,\lambda)\frac{\partial I(x_0,y_0,\lambda)}{\partial\alpha_i}
\label{eq:coupledresponse}
\end{equation}
since the instrumental PSF does not depend on the $\alpha_i$. We recognize the derivative on the RHS as the response function of the modeled spectrum used in most inversion methods employing a ``greedy'' algorithm \lf{\citep[see for instance][for a description]{Cormen:2001:IA:580470}} to minimize a merit function \citep{1977A&A....56..111L}. The above expression gives us the true response of the data, given the undegraded response of the spectrum. We use this below to formulate the correct response matrix for a \lf{2-D} field of point sources, degraded by a known PSF.

Although not strictly necessary, in the following we assume that the spatial part of the PSF does not depend on the wavelength so that we can write
$$
\phi(x,y,\lambda)=\varphi(x,y)\psi(\lambda).
$$
There are many \lf{situations} where this assumption is not valid, but \lf{it} allows us to carry out the convolution over the wavelength before considering the spatial coupling, which makes the implementation somewhat simpler. We intend to relax this restriction in future work.

\subsection{Levenberg-Marquardt minimization}
The Levenberg-Marquardt minimization procedure \citep{1944QoAM...2..164L,1963SoAM...11..431M} is used extensively in many areas where the minimum of a \lf{nonlinear} function needs to be located. It uses an adjustable mixture of a fully linearized solution and a steepest descent gradient search to locate the global minimum of a function of an arbitrary number of variables. \lf{It's} robust convergence properties \lf{make} it the method of choice for many atmospheric inversion codes\lf{,} and \lf{it} is also at the core of the SPINOR code.

The method is built on a linearization of the fit-quantity (the spectrum) around a ``best guess'' set of fit parameters (the atmospheric parameterization parameters). It requires \lf{computing} the derivative of each synthesized data point $y_i$ \lf{with respect} to each fit parameter $\alpha_i$, each of which composes one element of the so-called Jacobian matrix ${\bf J}$.

Although the calculation of ${\bf J}$ is generally possible only by using a finite-difference approximation with respect to the fit quantity, the specific form of radiative transfer equation (RTE) allows us to compute them by moving the derivatives inside the formal solution integral \citep{1977A&A....56..111L,1989ApJ...339.1093R,1992ApJ...398..375R} and evaluating the RTE using the derivatives of the opacity and emissivity with respect to the atmospheric quantities instead of the opacity and emissivity themselves. The resulting response functions (RFs) can be computed at a relatively \lf{modest} additional cost alongside the emergent spectrum and \lf{correspond to} the partial derivatives of the spectrum \lf{with respect} to the fit \lf{quantities}, required by the Levenberg-Marquardt minimization procedure.

Assuming the response functions are known, the solution ${\bf \alpha}=$ must describe the observed data ${\bf d}$
$$
{\bf J} {\bf \alpha}={\bf d}.
$$
For an approximate solution ${\bf \alpha^\prime}$
$$
{\bf J} {\bf \alpha^\prime}={\bf d^\prime}\neq {\bf d}.
$$
so that by using the linearity of {\bf J} we may write
\begin{equation}
{\bf J} ({\bf \alpha}-{\bf\alpha^\prime})={\bf J} {\bf \nu}={\bf d}-{\bf d^\prime}=\beta,
\end{equation}
where ${\bf \nu}$ is the linear correction for the set of atmospheric parameterization values $\alpha$ and $\beta$ is the difference between the observed data and the spectrum synthesized using ${\bf \alpha}$. Since we generally have more data points then fit parameters, our desire to invert ${\bf J}$ directly to obtain ${\bf \nu}$ is frustrated by ${\bf J}$ \lf{not being square}. The solution is to multiply both sides \lf{by} the transpose matrix ${\bf J}^{T}$, thus creating the pseudo-inverse problem
\begin{equation}
{\bf J}^{T}{\bf J} {\bf \nu}={\bf J}^{T} \beta,
\label{eq:direct}
\end{equation}
where \lf{multiplying} $\beta$ \lf{by} ${\bf J}^{T}$ can be seen as converting the \lf{overdetermined} but exact problem into a least square fitting problem. Although this linear system gives us the least\lf{-}square solution to the linearized problem, unfortunately, the original inversion problem is usually far from linear, so that direct application of this solution may be anything but an improvement. An additional damping \lf{constant} $\lambda_M$ is therefore added to (\ref{eq:direct}) to control the behavior. The result is the familiar Levenberg-Marquardt form
\begin{equation}
({\bf J}^{T}{\bf J}+\lambda_M {\rm\bf diag}[{\bf J}^{T}{\bf J}]) {\bf \nu}=\delta,
\end{equation}
with $\delta={\bf J}^{T} \beta$ and $\lambda_M$ the Marquardt damping \lf{constant}. The matrix on the RHS can now be inverted to solve for the correction vector ${\bf \nu}$. It is easy to see that for increasing values of the damping \lf{constant} $\lambda_M$, the fit matrix becomes more and more diagonal, while the corrections become smaller and smaller, thus approaching the steepest descent minimum search, which is known to be very slow and very vulnerable to finding a local minimum that is not the global minimum, but is guaranteed to lead to a decrease in the value of the merit function.

The convergence strategy is now clearly to keep the damping parameter as small as possible, in order to avoid the slow convergence \lf{properties} of the steepest descent method, but large enough not to overstep the minimum so far as to increase the merit function. Ideally, the initial steps should only \lf{be concerned} with the large\lf{-}scale behavior of the minimization function; \lf{small-scale details should only come into focus once the dominant features} have already been fairly well fitted. Unfortunately, as with all blind search algorithms, no guarantee exists that the minimum that is located is the lowest minimum.

\subsection{Spatial coupling}
Now that we understand the way in which the Levenberg-Marquardt method optimizes the merit function, we are ready to consider the spatial coupling between the pixels in an actual dataset.

There are at least two important ways in which the spectra in adjacent pixels depend directly on each other: horizontal radiative transfer effects and instrumental scattering and diffraction effects. The first involves the considerable task of solving the RTE in non-LTE and in full 3-D and computing the response functions for all image elements and all fit parameters in these image elements. Although it is not \lf{clear a priori} that this effect \lf{does not have to be} consider in general, we assume here that there are at least some photospheric lines for which the assumption of LTE is approximately valid \lf{so} this effect can be neglected. The second way is easily seen to constitute a much simpler problem than the first, since the local derivative of the spectrum to the atmospheric fit parameters is simply spread to the neighboring pixels according to the PSF.


Taking a closer look at the Jacobian matrix of a spectro-polarimetric image makes \lf{it} clear that for spatially uncoupled inversions, without a PSF, the Jacobian is simply an assembly of uncoupled sub-Jacobians:
\begin{equation}
 {\bf J}=
   \left(\begin{array}{c c c c c}
   {\bf J}_{1,1}& & \begin{array}{c c c}.&.&.\end{array} & & 0 \\
    &{\bf J}_{1,2}& & & \\
   \begin{array}{c}.\\.\\.\end{array}& &\begin{array}{c c c}.&&\\&.&\\&&.\end{array}& & \begin{array}{c}.\\.\\.\end{array} \\
    & & &{\bf J}_{n,m-1}& \\
   0 & &\begin{array}{c c c}.&.&.\end{array}& &{\bf J}_{n,m}\\
   \end{array}\right)
\end{equation}
so that we simply have
\begin{equation}
 {\bf J}^T=
   \left(\begin{array}{c c c c c}
   {\bf J}_{1,1}^T& & \begin{array}{c c c}.&.&.\end{array} & & 0 \\
    &{\bf J}_{1,2}^T& & & \\
   \begin{array}{c}.\\.\\.\end{array}& &\begin{array}{c c c}.&&\\&.&\\&&.\end{array}& & \begin{array}{c}.\\.\\.\end{array} \\
    & & &{\bf J}_{n,m-1}^T& \\
   0 & &\begin{array}{c c c}.&.&.\end{array}& &{\bf J}_{n,m}^T\\
   \end{array}\right)
\end{equation}
and the required fit matrix ${\bf J}^{T}{\bf J}$ clearly remains block-diagonal and can be solved efficiently block-by-block.

For spatially coupled inversions with a {\em uniform} PSF $\varphi(x,y)$ that only depends on the spatial coordinates, the Jacobian can be derived by making use of (\ref{eq:coupledresponse}) and can still be written in a blocked form:
{\small
$$
 {\bf J}=
   \left(\begin{array}{c c c c c}
   \varphi_{0,0}{\bf J}_{1,1}&\varphi_{0,-1}{\bf J}_{1,2}& \begin{array}{c c c}.&.&.\end{array} &\varphi_{1-n,2-m}{\bf J}_{n,m-1}&\varphi_{1-n,1-m}{\bf J}_{n,m}\\
   \varphi_{0,1}{\bf J}_{1,1}&\varphi_{0,0}{\bf J}_{1,2}& &\varphi_{1-n,3-m}{\bf J}_{n,m-1}&\varphi_{1-n,2-m}{\bf J}_{n,m}\\
   \begin{array}{c}.\\.\\.\end{array}& &\begin{array}{c c c}.&&\\&.&\\&&.\end{array}& & \begin{array}{c}.\\.\\.\end{array} \\
   \varphi_{n-1,m-2}{\bf J}_{1,1}&\varphi_{n-1,m-3}{\bf J}_{1,2}& &\varphi_{0,0}{\bf J}_{n,m-1}&\varphi_{0,-1}{\bf J}_{n,m}\\
   \varphi_{n-1,m-1}{\bf J}_{1,1}&\varphi_{n-1,m-2}{\bf J}_{1,2}&\begin{array}{c c c}.&.&.\end{array}&\varphi_{0,1}{\bf J}_{n,m-1}&\varphi_{0,0}{\bf J}_{n,m}\\
   \end{array}\right),
$$
}\\
but is clearly no longer block-diagonal. In practice, the extent over which $\varphi$ does not vanish is much smaller than the computational domain, resulting in a relatively sparse system. The transpose is readily written down as
{\small
$$
 {\bf J}^T=
   \left(\begin{array}{c c c c c}
   \varphi_{0,0}{\bf J}_{1,1}^T&\varphi_{0,1}{\bf J}_{1,1}^T& \begin{array}{c c c}.&.&.\end{array} &\varphi_{n-1,m-2}{\bf J}_{1,1}^T&\varphi_{n-1,m-1}{\bf J}_{1,1}^T\\
   \varphi_{0,-1}{\bf J}_{1,2}^T&\varphi_{0,0}{\bf J}_{1,2}^T& &\varphi_{n-1,m-3}{\bf J}_{1,2}^T&\varphi_{n-1,m-2}{\bf J}_{1,2}^T\\
   \begin{array}{c}.\\.\\.\end{array}& &\begin{array}{c c c}.&&\\&.&\\&&.\end{array}& & \begin{array}{c}.\\.\\.\end{array} \\
   \varphi_{1-n,2-m}{\bf J}_{n,m-1}^T&\varphi_{1-n,3-m}{\bf J}_{n,m-1}^T& &\varphi_{0,0}{\bf J}_{n,m-1}^T&\varphi_{0,1}{\bf J}_{n,m-1}^T\\
   \varphi_{1-n,1-m}{\bf J}_{n,m}^T&\varphi_{1-n,2-m}{\bf J}_{n,m}^T&\begin{array}{c c c}.&.&.\end{array}&\varphi_{0,-1}{\bf J}_{n,m}^T&\varphi_{0,0}{\bf J}_{n,m}^T\\
   \end{array}\right),
$$
}\\
which, using $ \varphi*\varphi=Y$, allows us to write the desired matrix as
\begin{equation}
{\bf J}^T{\bf J}=
   \left(\begin{array}{c c c}
   Y_{0,0}{\bf J}_{1,1}^T{\bf J}_{1,1}& \begin{array}{c c c}.&.&.\end{array} &Y_{1-n,1-m}{\bf J}_{1,1}^T{\bf J}_{n,m}\\
   Y_{0,1}{\bf J}_{1,2}^T{\bf J}_{1,1}& &Y_{1-n,2-m}{\bf J}_{1,2}^T{\bf J}_{n,m}\\
   \begin{array}{c}.\\.\\.\end{array}&\begin{array}{c c c}.&&\\&.&\\&&.\end{array}& \begin{array}{c}.\\.\\.\end{array} \\
   
   Y_{n-1,m-2}{\bf J}_{n,m-1}^T{\bf J}_{1,1}& &Y_{0,-1}{\bf J}_{n,m-1}^T{\bf J}_{n,m}\\
   Y_{n-1,m-1}{\bf J}_{n,m}^T{\bf J}_{1,1}&\begin{array}{c c c}.&.&.\end{array}&Y_{0,0}{\bf J}_{n,m}^T{\bf J}_{n,m}\\
   \end{array}\right)
\end{equation}
which contains the matrix product of all ${\bf J}^T_{i,j}$ with all ${\bf J}_{x,y}$, weighted with the autocorrelation function of the PSF centered on the location of the ${\bf J}_{x,y}$. 

If \lf{the PSF is not uniform}, it is straightforward to replace the \lf{coefficients} $\varphi_{i,j}$ with appropriate vectors containing the elements of the appropriate PSF. Although the computational overhead of this is probably not prohibitive, we leave this \lf{to be developed in} a future paper.

The Herculean effort of directly inverting this linear system, containing  $(n_a\times n_s\times n_x\times n_y)^2$ elements, must not only be considered well out of reach of any computing resource likely to be available in the near future, if it were accomplished by some means\lf{,} it \lf{is} very likely that the large number of operations needed for \lf{computing} the solution vector would lead to \lf{acumulation of} unacceptable numerical errors, since the inverse matrix would likely not be sparse.

However, since the forward problem {\em is} sparse, we are able to accurately evaluate
\begin{equation}
 ({\bf J}^{T}{\bf J}-\lambda\,{\rm\bf diag}[{\bf J}^{T}{\bf J}]) {\bf x}_n={\bf A} {\bf x_n}=\delta_n
\end{equation}
and compute the defect ${\bf d}_n$ in the current estimate ${\bf x}_n$ of the correction vector ${\bf \nu}$
\begin{equation}
  {\bf d}_n=\delta-\delta_n
 \label{eq:resideq}
\end{equation}
in a reasonable amount of time. We now approximate ${\bf A}$ with a sufficiently accurate approximation, ${\bf A}^*$, \lf{which} is easy to invert but captures the essence of ${\bf A}$. The solution of
\begin{equation}
 {\bf A}^* {\bf \zeta_n}={\bf d_n}
 \label{eq:approxeq}
\end{equation}
can now be used to improve the estimate
$$
{\bf x}_{n+1}={\bf x}_{n} + \zeta_{n},
$$
repeatedly, until $\mid {\bf d}_n\mid^2<\epsilon$, with $\epsilon$ a suitably small, positive number.
This process converges rapidly if \lf{an} accurate approximation of ${\bf A}$ is used.

\subsection{Approximate operator}
\label{sec:shortcut}
In the process outlined above, an approximation to the full matrix ${\bf A}={\bf J}^{T}{\bf J}-\lambda_M\, {\rm\bf diag}[{\bf J}^{T}{\bf J}]$ is required that captures the basic properties of the full operator but is both cheap to invert and has a sparse inverse. The block diagonal approximation is an obvious candidate, \lf{because it} works well for a compact PSF, but requires a large number of iterations to converge when a more extended PSF is used.

\begin{figure}[htb]
\includegraphics[height=2.90cm]{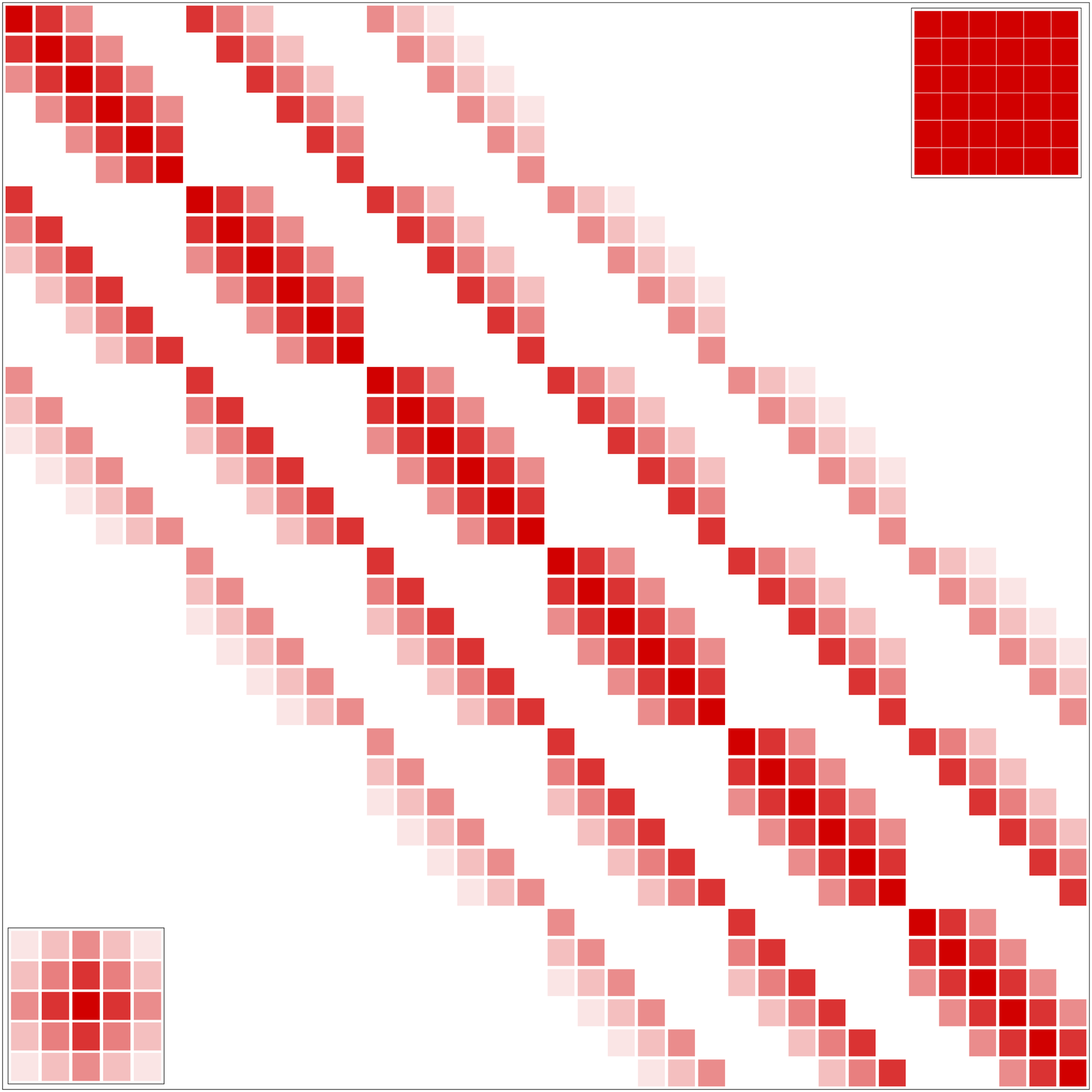}
\includegraphics[height=2.90cm]{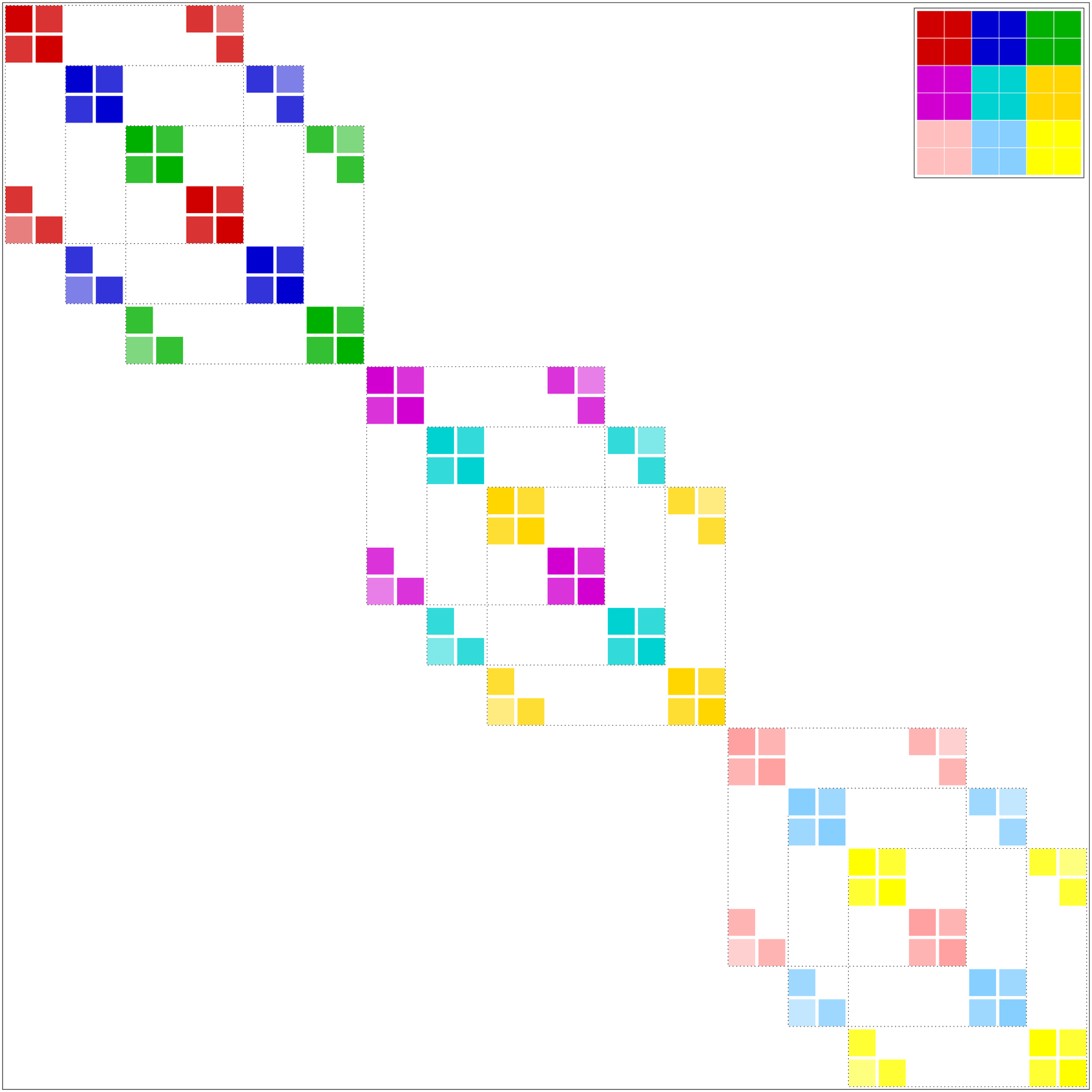}
\includegraphics[height=2.90cm]{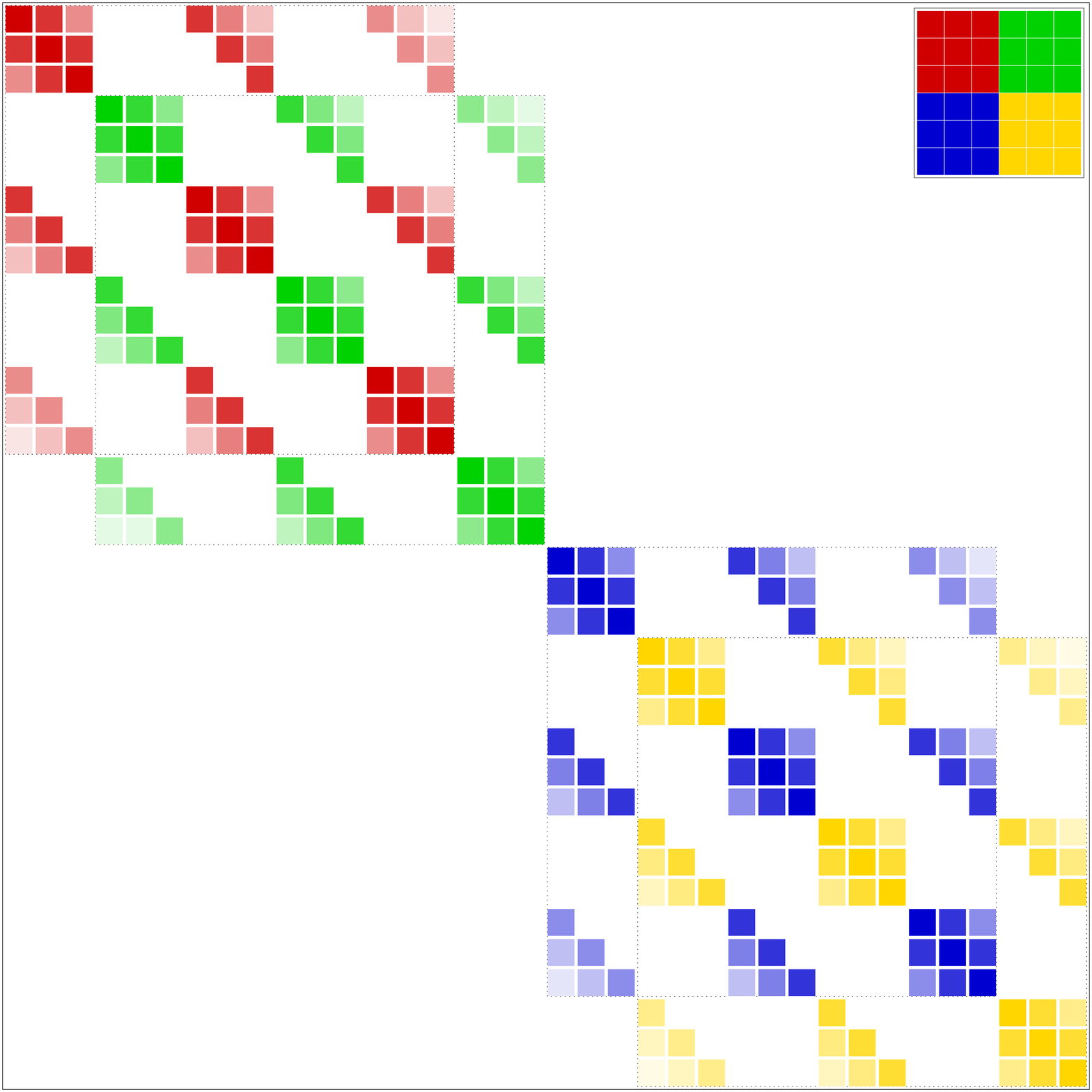}
\caption{Full matrix for a grid of 6$\times$6 pixels ({\bf left}), approximated by 9 blocks of $2\times 2$ pixels ({\bf middle}) and 4 blocks of $3\times 3$ pixels ({\bf right}). The cost increases rapidly with the number of pixels $N$, as the inversion of the resulting matrix scales $\propto N^3$. In the upper right corner, the spatial distribution of the pixels is indicated \lf{by} the color of the block they belong to. The full matrix has in addition the cross-correlation function, ${\bf Y}$, inserted in the lower left corner.}
\label{fig:fullmat}
\end{figure}

An extended-block variant of this method was also tried, where a spatially connected region was selected and inverted explicitly, as illustrated in Figure~\ref{fig:fullmat}. This method generally improves the convergence characteristics considerably, but comes at the increased 
cost of inverting a larger part of ${\bf A}$ explicitly. Apart from the increased cost of the inversion, this approach suffers from the limitation that when the size of the blocks becomes too large, the numerical errors in the inverted block matrix can become large enough for the iterative approximation to fail to converge completely. It is therefore important to find the optimum balance between the cost of explicit inversion and the cost of applying a larger number of iterations, which generally \lf{depends on the problem}.

Other variations on the extended block method can be made, when more \lf{nonlocal} behavior needs to be captured adequately.
\begin{figure}[htb]
\includegraphics[height=4.40cm]{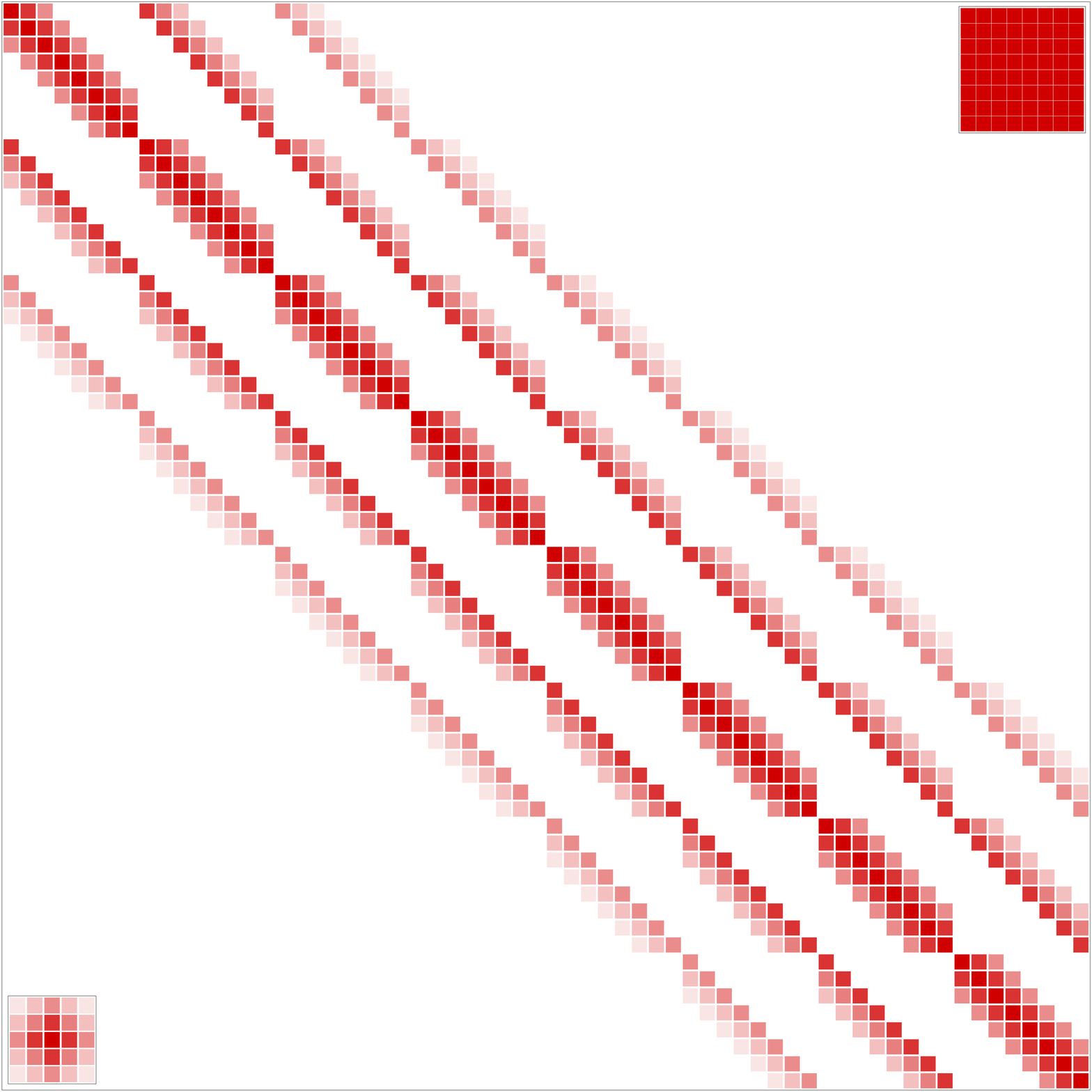}
\includegraphics[height=4.40cm]{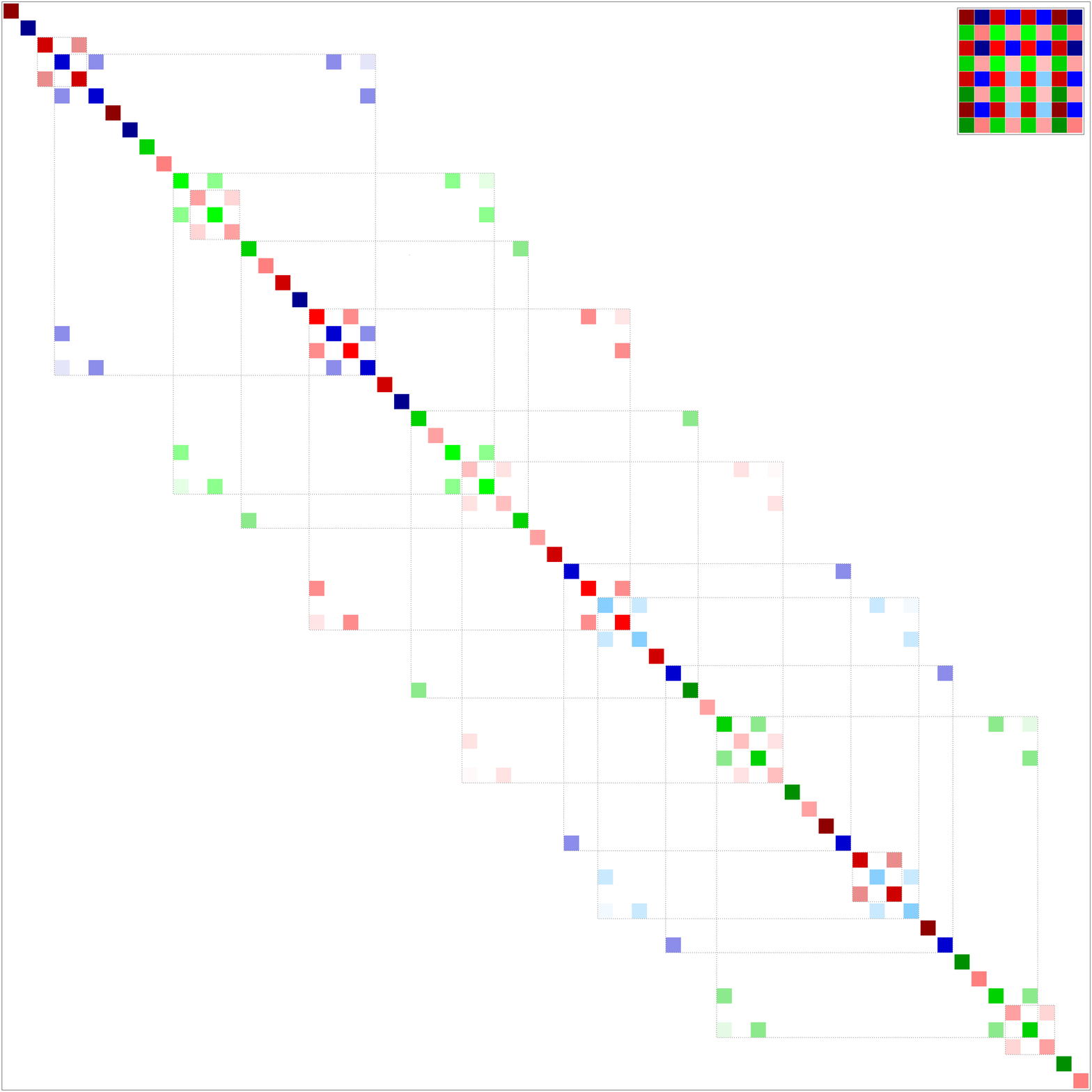}
\caption{Full matrix for a grid of 8$\times$8 pixels ({\bf left}), approximated using a staggered configuration of $2\times 2$ pixel blocks ({\bf right}). The approximated matrix has more than 9 blocks, due to boundary effects, compared with only 9 above, but many of the blocks are smaller. The insets are as in Figure~\ref{fig:fullmat}}
\label{fig:fullmatstag}
\end{figure}
The staggered block method from Figure~\ref{fig:fullmatstag} can be useful when interaction over long distances is important, such as for significantly oversampled data. Intermittent application of several methods is possible, but the need to recompute the inverse of the approximate matrix generally outweighs the \lf{possible} gain in convergence.

Computational optimization using explicit storage of all nonzero elements of $({\bf J}^{T}{\bf J}-\lambda_M\, {\rm\bf diag}[{\bf J}^{T}{\bf J}])$ and the approximation of the inverse, in combination with \lf{multithreading} of the iterative solver, reduces the additional time overhead of this method to only $\approx$30\% of the cost of synthesizing the Stokes profiles and response functions \fix{with the SPINOR code}, for critically sampled data and \lf{with} observational data containing $\sim$100 spectral points. This overhead is expected to decrease further when the amount of spectral points increases.

Depending on the approximate operator that is used, the convergence of the inversion of ${\bf A}$ \lf{typically requires} on the order of 100 iterations to reach \lff{an} error in the defect of 10\%, as shown in Figure~\ref{fig:da2d}. This takes significantly less time than the computation of ${\bf A}$ and ${\bf A}^*$ and so does not significantly contribute to the overall execution time, as long as the inverted \lf{submatrices} can be stored in RAM.

Unfortunately, the need to update the correction vector and recompute the defect, requires knowledge of the current estimate across the extent of the ${\bf Y}$ functions, resulting in the need for a significant amount of communication when a distributed parallelization strategy is desired. For this reason, at present, only an efficient \lf{multithreaded} algorithm exists, requiring a shared memory environment.

\lf{In some cases, additional shortcuts can be taken to reduce the work requirements. Such shortcuts, however,} do not always result in unconditional convergence, as is the case for the complete treatment as outlined above. One particularly useful approximation in the case of a PSF with very extended wings is to calculate the defect $\beta$ explicitly using the complete PSF, \lff{while calculating} the Levenberg-Marquardt correction ${\bf\nu}$ using only the central part of the PSF. This results in good convergence when the value of the PSF integrated over the extended part of the PSF does not exceed the \lf{value of the PSF integrated across the central part}. If this \lf{condition is not met}, an exponentially growing correction is obtained and the inversion process will diverge.

\subsection{Inversion strategy}
Although we can evaluate the global problem, it is not self-evident that this is the optimal way to proceed. If we consider the pixel-by-pixel approach as completely uncoupled and a very extended PSF as a global inversion problem, it is clear that in general all levels of coupling will be encountered. For the calculations presented in this paper, only the global problem \lf{has been considered;} however, it should be kept in mind that the global approach may not be the optimal choice in all cases.

\begin{figure}[htb]
\includegraphics[height=3.10cm,bb=0 0 254 90]{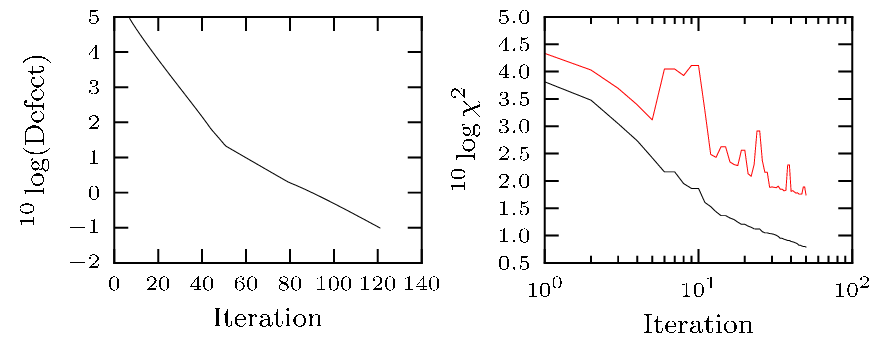}
\caption{Convergence properties of the iterative step determination and the global inversion {\bf left}: error in the defect as a function of iteration number, {\bf right}: average (black) and maximum (red) $\chi^2$ as a function of iteration number.}
\label{fig:da2d}
\end{figure}

The value of the merit function, $\chi^2$, is plotted in Figure~\ref{fig:da2d} for the inversion of an MHD simulation as a function of the 
iteration number. The convergence rate can be strongly influenced by employing additional smoothing at regular intervals, which can effectively push the solution out of local minima by damping out large local differences \lf{that typically result} from the cancellation of errors by other errors with \lf{the} opposite sign.

A returning issue when converging oversampled data is the development of a large number of local minima, frequently involving spatial frequencies in the solution exceeding the diffraction limit. One of the more effective ways to reduce the probability of getting stuck in one of these minima is to initially limit the corrections applied to the solution to the lower frequencies \lf{and to relax} this limitation as the inversion progresses.

Another effective way to avoid a slow and cumbersome descent to the global minimum is to introduce a significant amount of noise in the data in the first stages of the inversion. The noise effectively hides the less significant spectral features from the view of the inversion algorithm, leading to a rapid first descent phase. After the minimum has been approached with sufficient accuracy, the noise is gradually reduced, allowing for the details to enter the solution. This method is fairly effective in avoiding getting captured by outlying local minima, although it is not very effective in avoiding false minima near the global minimum. 

\section{Simulations}
\label{sec:simulations}

\begin{figure*}[!ht]
\includegraphics[width=\textwidth, bb=0 0 510 204, clip]{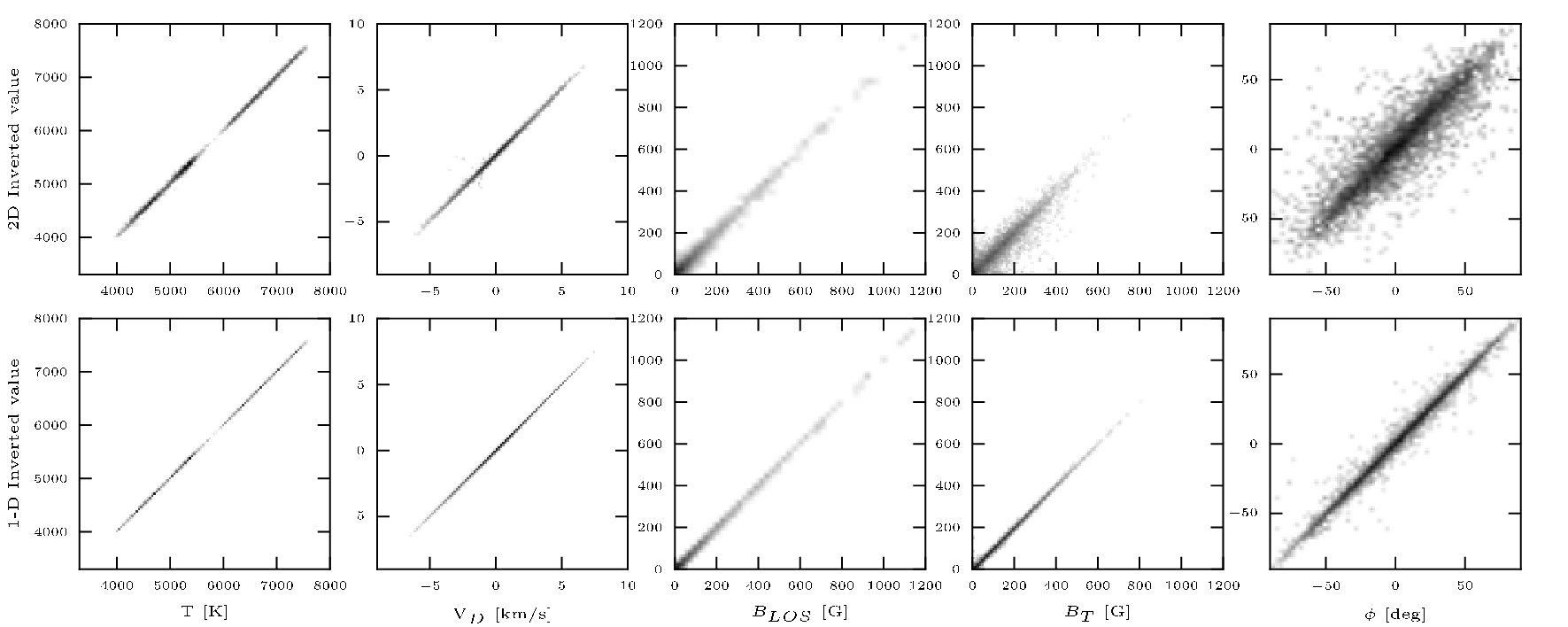}
\caption{2-D histograms of inverted atmospheric parameters (vertical) as a function of input atmospheric parameters (horizontal), for a coupled inversion of simulated degraded 50cm telescope data (top row) and an uncoupled inversion of identical but {\em undegraded} simulated data (bottom row), for 5 atmospheric parameters across all heights in the atmosphere.}
\label{fig:critical}
\end{figure*}
\lf{To} evaluate and optimize the performance of the spatially coupled inversion problem, it is necessary to know what the value of a particular atmospheric parameter really is, something that can only be done in the context of simulated atmospheres. The atmosphere chosen for the tests below is a 288x288x100 point MURaM MHD simulation \citep{2005A&A...429..335V} of the quiet Sun with a mean field of around 100G and a spatial resolution of 20.8km (0.029"). The grid size allows for convenient binning in a number of coarser grid sizes, which we make use of in the tests carried out below.
\begin{figure}[htb]
\hspace*{2.00cm}\includegraphics[height=4.50cm]{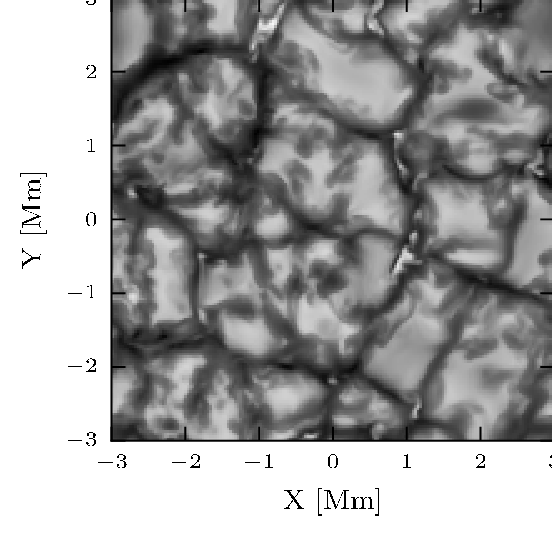}
\caption{The continuum intensity of the employed MURaM quiet Sun simulation.}
\label{fig:icont}
\end{figure}

The verification procedure necessarily remains strictly limited to validation and verification of the method under inherently ideal circumstances, since the forward and the backward \lf{problems} use the same physics, a situation that is not likely to occur in the case of real observations. The effects of errors in the physics will be considered inherent to all inversions and are left for a later discussion.

The starting point of our calculations is an MHD cube, pictured in Figure~\ref{fig:icont}\lf{,} that provides the relevant atmospheric quantities (temperature, pressure, velocity and magnetic field) as a function of \lf{three} spatial coordinates. An unambiguous comparison of such a fully stratified atmosphere with an inverted simplified atmosphere is a complicated task by itself and one that is not unique to the coupled inversion problem.

To steer clear of a complicated discussion on this topic, the atmosphere was first used to synthesize the emergent spectra, which were then inverted using simple \lf{three} node atmospheres, with the nodes at the same heights as they will be for the inversions. The resulting atmospheres are only approximations to the original ones, but they have the advantage that they can be represented exactly by the inversion code.

The fitted atmospheres are then used to generate a second data cube of emergent spectra, with the notable difference that these spectra can be \lf{reproduced exactly} by the simplified \lf{three}-node atmospheres used by the inversion code, so that it should be possible to invert the spectra perfectly to the input atmospheres used to generate them, provided the mapping of the atmosphere to the emergent spectra is unique.

\subsection{Ideal atmospheres}
\begin{figure*}[!ht]
\includegraphics[width=\textwidth, bb=0 0 510 255, clip]{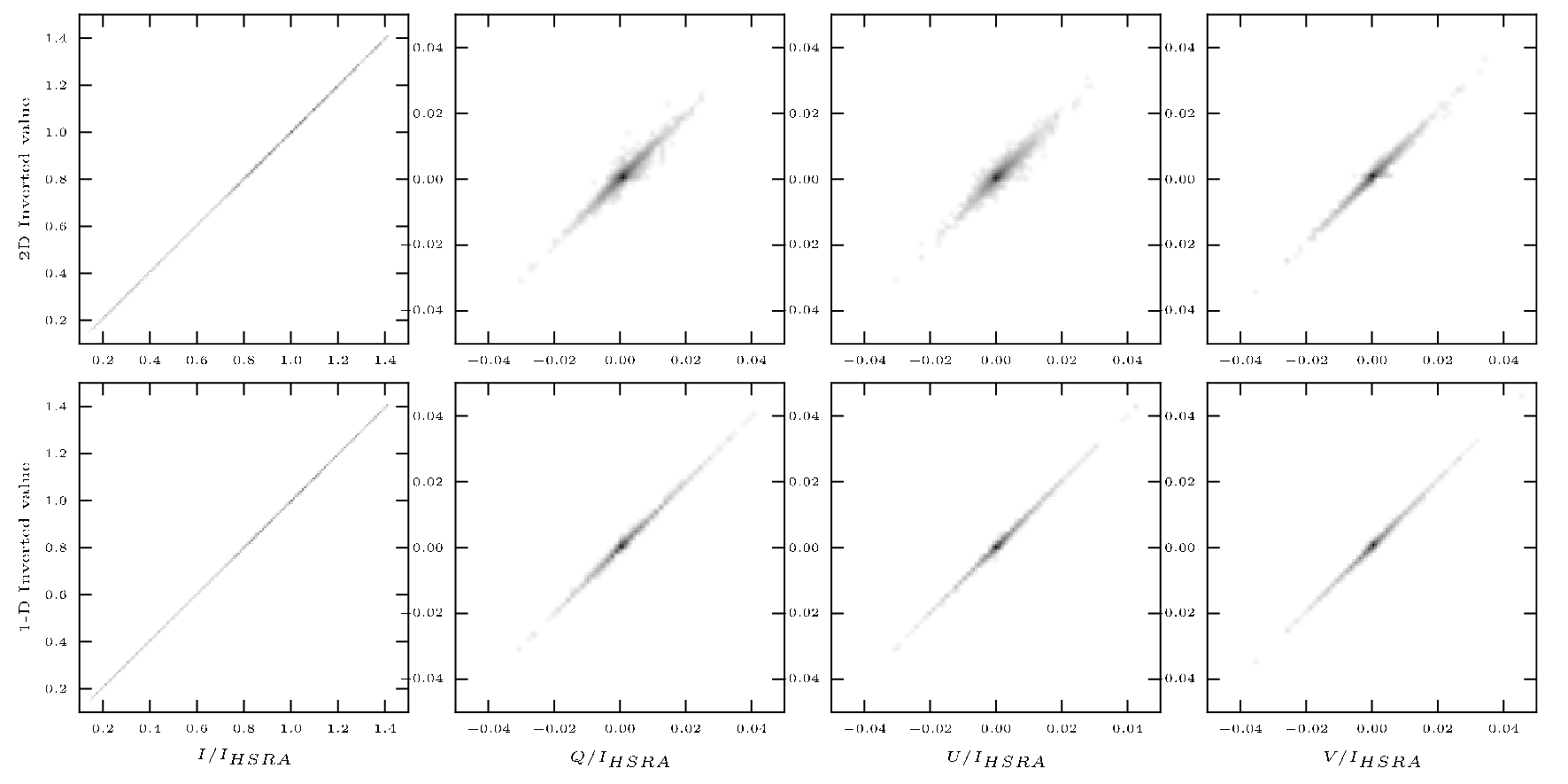}
\caption{\lf{2-D} histograms of inverted Stokes parameters (vertical) as a function of input Stokes parameters (horizontal), for a 2-D inversion of degraded 50cm telescope data (top row) and a 1-D inversion of identical but undegraded data (bottom row).}
\label{fig:critical_iquv}
\end{figure*}
As motivated \lf{and explained} above, we attempt to validate the method under the simplest possible circumstances, where the profiles are generated from a known solution that can be \lf{reproduced exactly} by the inversion code. \lf{To} achieve this, the \lf{three}-node atmospheres\lf{,} fitted to the synthesized spectra as described above, were first binned to the resolution of the artificial dataset we want to generate, {\em before} the spectra to be degraded and inverted are synthesized from them. 

If the inversion problem is \lf{well-posed}, this simplification should allow the inversion code to recover the binned atmosphere with arbitrary precision in the absence of spectral and spatial degradation. Since this is generally not the case for all pixels in the field of view (a few spectra will always be difficult to invert), to evaluate the properties of the coupled inversion as independently of the properties of the inversion procedure as possible, we focus on the differences between the pixel-by-pixel inversion of the undegraded data and the coupled inversion of degraded data.

\subsubsection{Pixel-by-pixel inversions}
\label{sec:pbpi} After a first verification that the atmosphere used to generate the spectra gives a perfect match to the artificial data, a pixel-by-pixel inversion was carried out of the whole FOV, to test the inherent invertibility of the atmospheres using the algorithm used by the inversion code. Interestingly, the results indicate that, although the code should be able to perfectly reproduce the spectra, it does not succeed in doing so for approximately 1\% of the pixels in the FOV. 

Repeated  ``nudges'' of the atmospheric parameters in random directions in parameter space improve the overall success rate, but in a number of pixels, the solution is evidently difficult to locate using the minimization procedure employed by the code, so that significant differences between the artificial observation and the inverted spectrum remain. However, where the inversion code indicates a good fit, the atmospheric parameters are accurately retrieved, suggesting that the inversion problem is not \lf{underdetermined} in this simplified situation.

The fraction of the pixels that does not produce an accurate inverse is approximately independent of the resolution, which is somewhat surprising, since although the atmospheres are binned to the resolution of the data, so that no unresolved structure can be present, the profiles tend to exhibit more extreme behavior at higher resolution. Figure~\ref{fig:critical} shows the retrieved inverted values as a function of the input values. The temperature and Doppler velocity appear to be very \lf{well-determined} in all layers of the atmosphere, but the magnetic field strength and orientation show minor scatter around the diagonal, indicating that there are limits to the invertibility of the problem.

Figure~\ref{fig:critical_iquv} shows the recovered Stokes parameters as a function of the input Stokes parameter. The agreement is very good in all Stokes parameters, indicating that the minor scatter observed in the recovered magnetic field parameters is due to [near] degenerate behavior in these parameters, possibly \lf{owing} to an inherently low response.

\begin{figure*}[!ht]
\includegraphics[width=\textwidth, bb=0 0 525 311, clip]{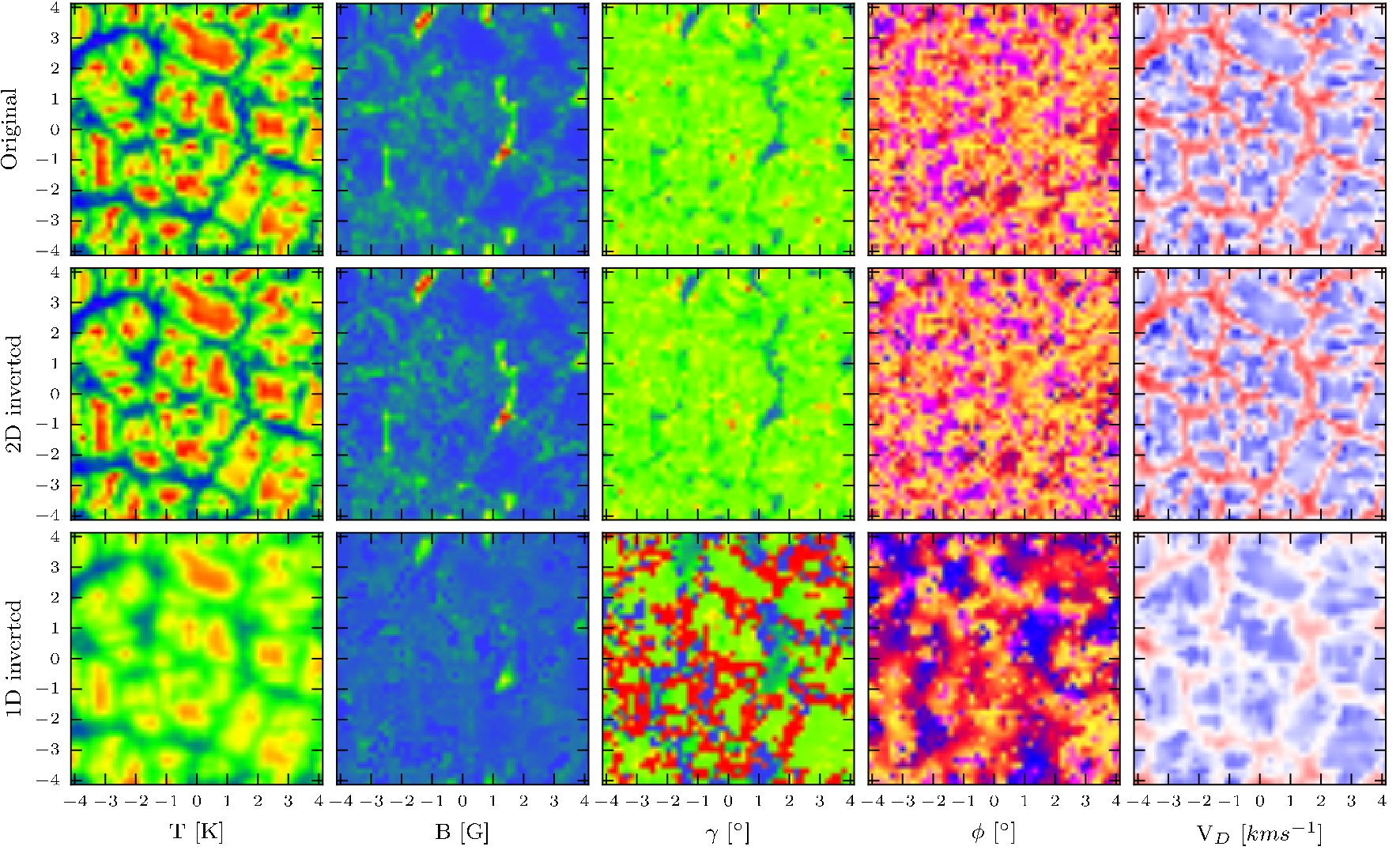}
\caption{Visual comparison of a selection of inverted parameters from spatially degraded spectra with the original data (top) for a spatially coupled inversion (middle) and a spatially uncoupled inversion (bottom). Shown are the parameter maps at the optical depths where they are most accurately determined. From left to right: Temperature at $\tau=1$, magnetic field strength, inclination angle and azimuth angle at $\tau=0.16$\lf{,} and the line-of-sight velocity at $\tau=1$. The color schemes are optimized for each quantity and have an identical but otherwise arbitrary scale for the original and inverted parameters. Tick marks indicate the spatial scale in \lf{arcseconds}.
Clearly visible is the difference in the accuracy of the retrieved parameters when the spatial smearing by the telescope is not taken into account.}
\label{fig:overview_comp}
\end{figure*}
\subsubsection{Spatially coupled inversions}
\label{sec:sci} In the spatially coupled case, we must choose a PSF to degrade the spectral data with. With a large volume of available potential data in mind, we choose the central part of the PSF of the SOT \citep{2008SoPh..249..167T} on board the Hinode spacecraft \citep{2007SoPh..243....3K}, as indicated in Figure~\ref{fig:hinodepsf}, so we get a direct indication of the performance we can expect for such data.

The original resolution of the MHD cube, 0.029"/pixel, is approximately a factor 9 higher than the diffraction limit of 0.26" of the Hinode SOT at 6300\AA, so that a binning factor of 5 would be the closest approximation of critical sampling. Since this is not an integer factor of 288, the cube was binned by a factor 6 to $48\times 48$ pixels, corresponding to 0.174"/pixel, which is slightly \lf{undersampled} and in reasonable agreement with the Hinode spectro-polarimeter (SP) \citep{2001SPIE.4498...73L} pixel size of 0.16". In addition, a cube binned by a factor 3 to $96\times 96$ pixels was produced to investigate the response of the algorithm to oversampled data with unresolved fine structure.

\lf{Compared to a pixel-by-pixel inversion,} the spatially coupled inversions clearly require an increased amount of human intervention, \lf{since there are so many} coupled fit parameters that there is a significantly \lf{increased} probability of finding a local minimum that is not the global minimum. Identifying problematic pixels is difficult, because the only measure of the quality of the fit, the difference between the convolved profiles and the data, is the result of an ensemble of profiles, so that it is not possible to rate the contribution to the fit quality of one particular profile. A ``nudge'' of all fit parameters in a random direction is not necessarily helpful, since many pixels may have \lf{already converged} and a random step in that solution may lead to improved convergence in one place, but not in another.

For slightly \lf{undersampled} data, the method does not seem to have much more trouble recovering the solution than in the case of a \lf{1-D} inversion in the absence of any spatial degradation for the majority of the atmospheric parameters. Figure~\ref{fig:critical} shows the retrieved atmospheric quantities as a function of the input atmospheric quantities used to generate the undegraded spectra. Although the scatter around the diagonal is somewhat larger than for the pixel-by-pixel inversions of spatially undegraded data, when we keep in mind that the histograms in Figure~\ref{fig:critical} \lf{have a logarithmic gray scale}, we may conclude that on a linear scale, the only parameters with significantly increased scatter are the horizontal components of the magnetic field.

This is also clearly visible in the inverted Stokes spectra shown in Figure~\ref{fig:critical_iquv}, where the recovered Stokes Q and U values show the \lf{greatest} deviation from their real value. Since the value of the merit function does not vanish for this solution, we can state with confidence that despite the human intervention, the inversion code found a local minimum that is not the global minimum, which is the most likely cause for the observed errors.

As a final illustration, Figure~\ref{fig:overview_comp} shows the maps of a selection of the original parameters used to calculate the Stokes profiles, together with the inverted result. To emphasize the importance of correctly treating the spatial degradation, the uncoupled (1D) inversion result of the spatially degraded data is shown. Even though the parameters are shown at the optical depth where they are most accurately constrained by the data, the loss of accuracy caused by neglecting the spatial smearing of the data is clearly visible in the inverted results. We will explore the differences between the various inversion strategies in more detail in a future paper specifically on this topic.

\subsection{The effect of noise}
\begin{figure*}[!ht]
\includegraphics[width=\textwidth, bb=0 0 510 510, clip]{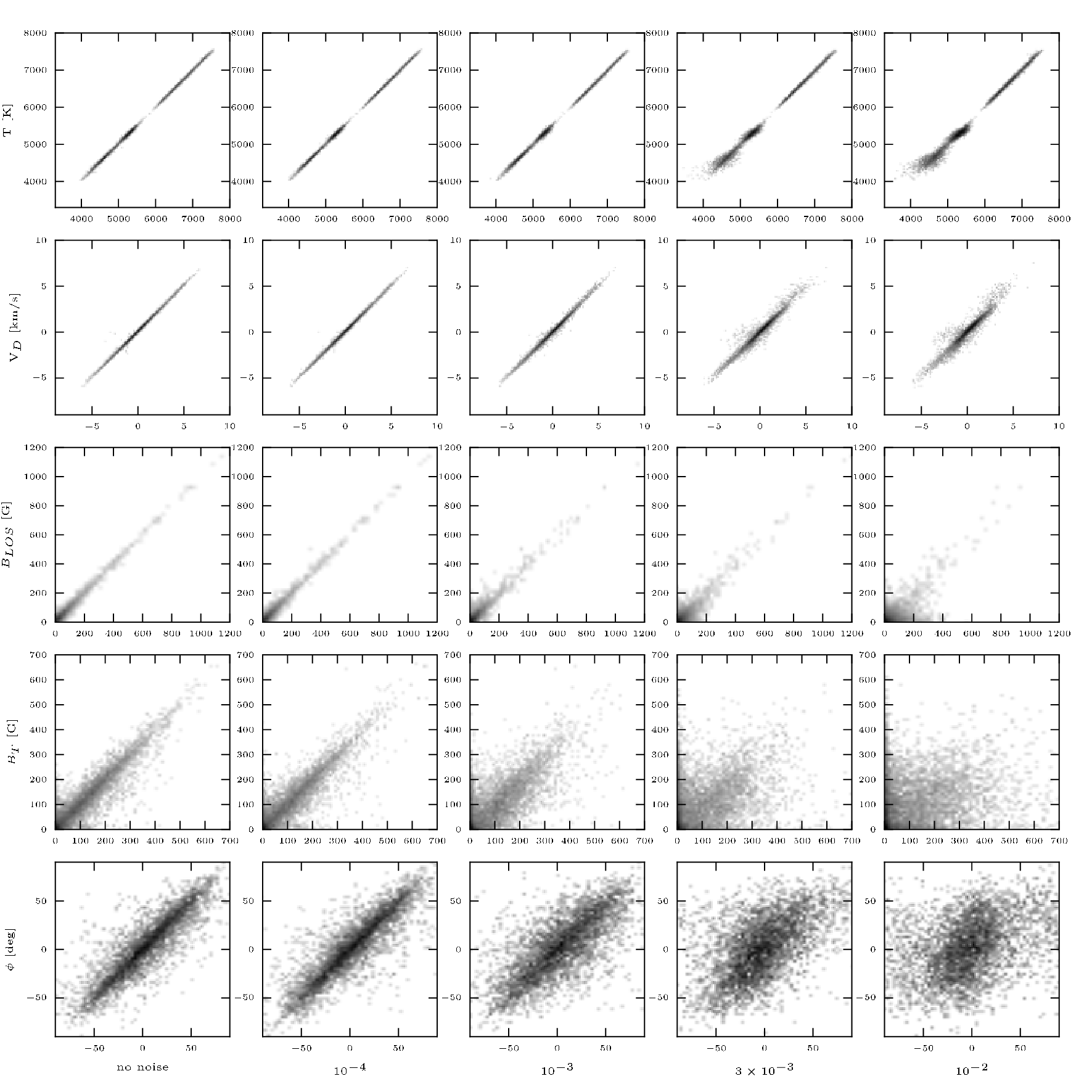}
\caption{Two-dimensional histograms of inverted atmospheric parameters (vertical) as a function of input atmospheric parameters (horizontal), for a 2-D inversion of degraded 50cm telescope data for noise levels of (left to right) 0, $10^{-4}$, $10^{-3}$, $3\times 10^{-3}$\lf{,} and $10^{-2}$ $I_c$, for 5 atmospheric parameters across all heights in the atmosphere.}
\label{fig:noise}
\end{figure*}
The effect of noise is one of the main problems to deal with when one is interested in the interpretation of real data. To investigate the effect of additive Gaussian noise on the spatially coupled inversions, \lf{a number of} synthetic observations were produced with noise levels \lf{that are typical for data} obtained with an efficient instrument and an integration time of several seconds.

The results of a repeat of the test from section \ref{sec:sci}, using data with varying levels of noise, is shown in Figure~\ref{fig:noise}. The critical level for reliable recovery of the atmospheric parameters appears to be $10^{-4}$ $I_c$, beyond which first the transversal (T) magnetic field properties rapidly degrade, followed by the longitudinal magnetic field and finally the temperature and line-of-sight (LOS) velocity. At a noise level of $10^{-2}$ $I_c$, practically all atmospheric parameters show large errors, suggesting that a level well below that should always be achieved for any atmospheric property to be reliably recovered. 

In particular\lf{,} the recovered temperature and line-of-sight velocity appear to \lf{only be} weakly affected by the noise. This is easily understood as these quantities may be inferred directly from the intensity, which is clearly the Stokes parameter least affected by the noise. In addition, the intensity in the continuum has a strong response to the temperature in the deepest layers, as well as a large number of spectral data points constraining that part of the profile.

The atmospheric quantities \lf{that only depend} on the Stokes Q, U\lf{,} and V parameters are significantly more sensitive to the noise, which is not only much more severe owing to the relatively low signal level of these data, but also to the negligible response of the majority of the (continuum) data points to these quantities. The final consequence of this is that for reliable recovery of the longitudinal component of the magnetic field in the quiet sun, a \lf{signal-to-noise} of $\sim  10^{-3}$ I$_c$ is required, for the orientation and field strength it must be significantly better.

\subsection{Unresolved structure}
\begin{figure*}[!ht]
\includegraphics[width=\textwidth, bb=0 0 510 204, clip]{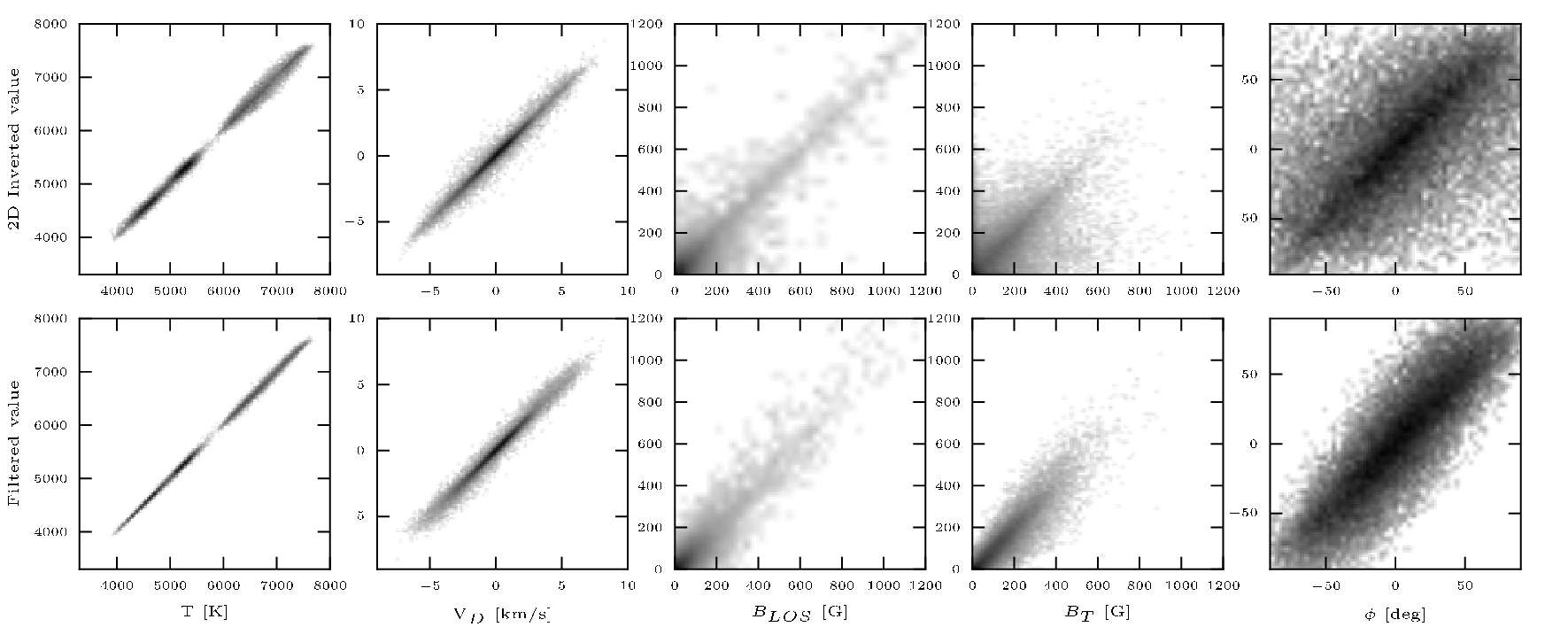}
\caption{Two-dimensional histograms of inverted atmospheric parameters (vertical) as a function of input atmospheric parameters (horizontal), for a 2-D inversion of degraded 50cm telescope data (top row) and a spatially filtered but otherwise undegraded atmosphere (bottom row), for 5 atmospheric parameters (horizontal) across all heights in the atmosphere.}
\label{fig:oversampled}
\end{figure*}

One of the main concerns in any inversion is the uniqueness of the solution, in other words\lf{,} is the problem invertible at all? Naively one might expect the spatially coupled fit to have problems here, since only the sum of a number of spectra is constrained, not each individual spectrum. The difficulty the code has in locating the global minimum solution suggests, however, that this is actually not the case, so that if there are multiple optimum solutions, clearly none of them is found.

In an attempt to test what happens to the solution in cases where this uniqueness should break down, an inversion was done on oversampled simulated data, a case where conventional convolution is known to no longer provide one-to-one mapping. To this end, a data cube was simulated as before, but this time at a resolution of 0.085"/pixel, and convolved with the Hinode PSF sampled at 0.085"/pixel, so that it is possible for the solution to contain information that is not constrained since it is no longer present in the inverted observation after convolution with the PSF.

Not unexpectedly, the inversion does not find \lf{as good a} solution as in the case of critically sampled data, even if just the value of the merit function is considered, but it shows no sign of instability or amplification of errors either. This suggests that the constraints provided by imposing a particular atmosphere and particular physics on the solution are able to stabilize the inversion sufficiently, even in oversampled cases.

Figure~\ref{fig:oversampled} shows the scatter of the inversion result as a function of the input atmosphere. Although the scatter is significantly increased over the critically sampled case, it is similar to the scatter of an appropriately spatially filtered atmosphere, shown in the bottom half of the same figure. The algorithm apparently does not recover much information beyond the diffraction limit, \lf{but} instead appears to recover an atmosphere that has all frequencies beyond the diffraction limit removed from it.

\subsection{Discussion}
The simulation results show that the evaluation of the merit function in convolved data space allows for accurate discrimination between profiles in the undegraded image to spatial scales up to the diffraction limit of the PSF, if the instrumental properties are known. The undegraded Stokes profiles are recovered with a minor increase in the absolute error, a behavior closely followed by the inverted atmospheric parameters. Clearly, the quantities that depend primarily on the Stokes I profiles are recovered more accurately than those depending on Stokes Q, U\lf{,} and V, even \lf{if only numerical noise is present}.

Interestingly, the inversion algorithm does not converge to the absolute minimum, where the merit function was verified to vanish. The difficulty is that whereas a pixel-by-pixel inversion allows the identification of poorly converged pixels, this is not possible in the spatially coupled case. 

The failure to find the absolute minimum within the space accessible by the algorithm suggests that if the solution is degenerate, the alternative degenerate solutions are as difficult to locate as the true solution, even in the case of oversampled data.

\section{Application to Hinode observations}
\label{sec:hinode}
Since the spectral dimension is crucial for this method to work and the simplification introduced by the \begin{figure}[htb]
\includegraphics[height=3.80cm]{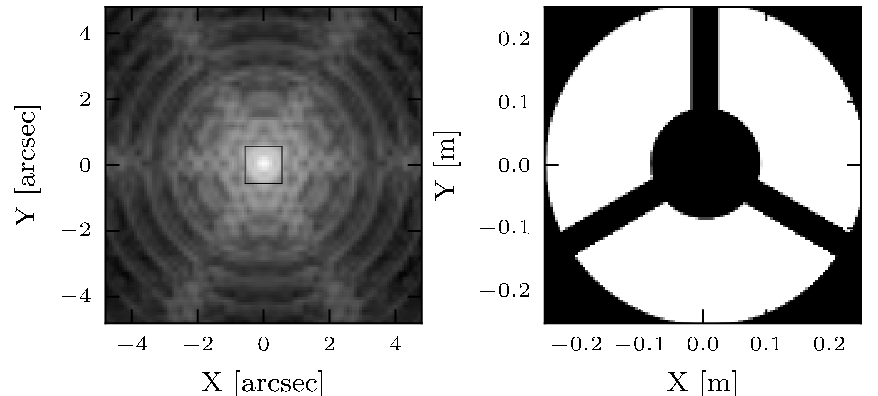}
\caption{{\bf left:} The Hinode PSF at 0.16" per pixel resolution on a logarithmic gray scale from $10^{-6}$ (black) to 1 (white). The black box is the explicit part used in the spatially coupled inversions. {\bf right:} Hinode SOT pupil function.}
\label{fig:hinodepsf}
\end{figure}
\begin{figure*}[!htb]
\includegraphics[height=17.00cm]{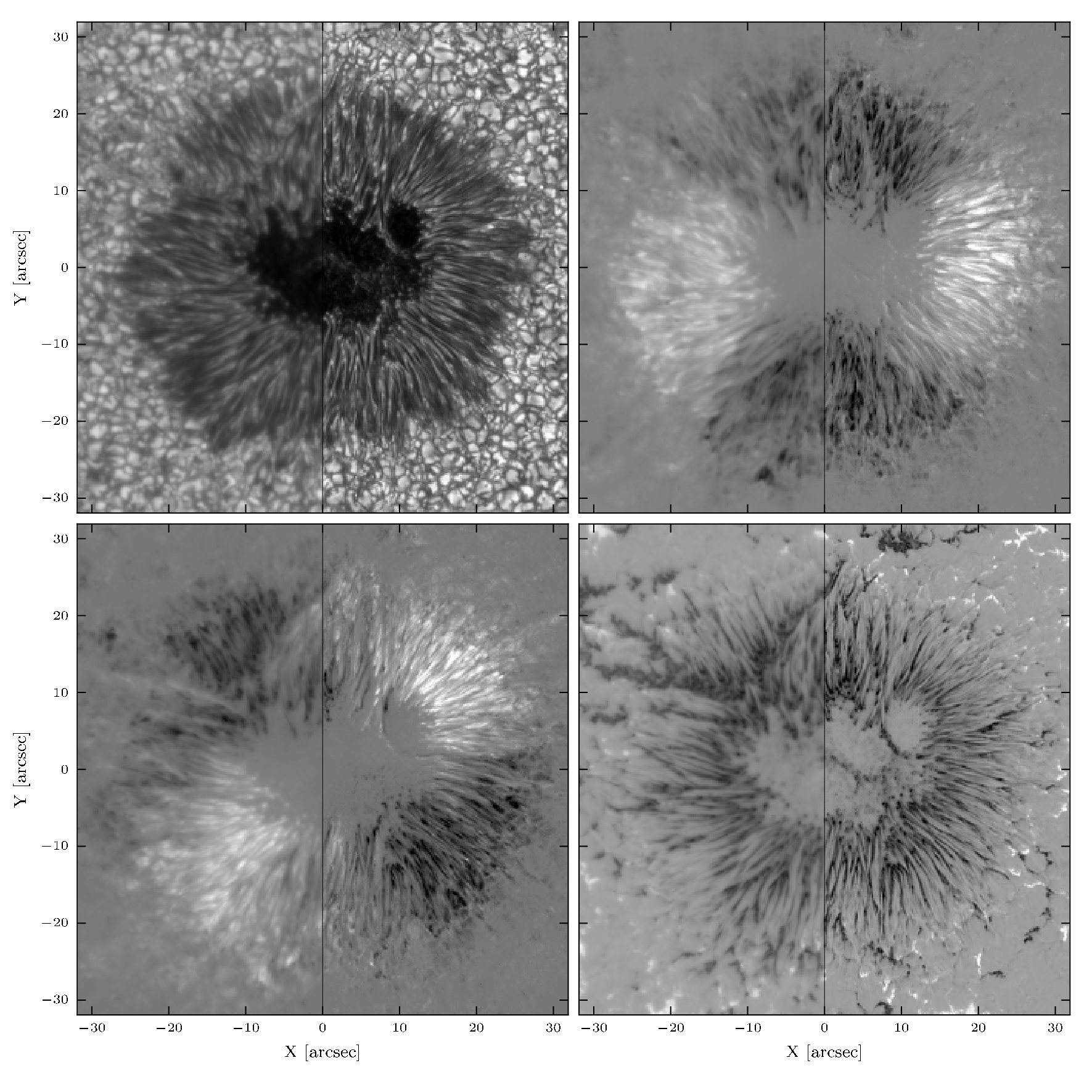}
\caption{Stokes images in the Fe\,I 6302.5\AA\ line+ 57 m\AA, before (left half) and after (right half) inversion for the Stokes parameters I (top left), Q(top right), U(bottom left) and V(bottom right).}
\label{fig:hinode2d}
\end{figure*}
assumption of a constant PSF is advantageous in keeping the computations tractable, the spectro-polarimetric scans of the Fe I lines at 6301.5\AA\ and 6302.5\AA\ provided by the Hinode spectro-polarimeter were used to test the performance of the coupled inversion method on real solar data.

The significant scanning time of this instrument compared to the evolution time scale of some solar structures gives reason to be critical of the result, however, the slight under-sampling of the data keeps the evolution effects of the solar image relatively small on the critical spatial scale of the PSF. Nonetheless, it is important to remember that accurate reproduction of the spectra may not be achieved in some cases.

In addition to this obvious shortcoming, we are dealing with a real solar atmospheric structure, with\lf{,} more likely than not\lf{,} a complicated depth stratification, which will not be accurately represented by the simple atmosphere imposed by the inversion code. \lf{Furthermore, although presumably very accurate, the assumption of LTE probably does not hold exactly and the atomic and molecular line data are not known perfectly.}

We nonetheless proceed by computing the PSF of the telescope. The PSF was computed from the Hinode pupil as specified in detail in the technical reference manual \citep{2008SoPh..249..197S}. As the focus position was found to differ from 0 by \cite{2008A&A...484L..17D}, \lf{a} 0.1 wave defocus was added, even though the focus position of this dataset was not accurately known. The PSF up to a radius of 4.8" is shown in Figure~\ref{fig:hinodepsf} on a logarithmic \lf{grayscale} from $10^{-6}$ to 1, showing a complicated and extended interference pattern with significantly enhanced radial structures produced by the triangular spider.

To calculate a 2D inversion using the whole PSF, an unreasonably large amount of memory and computing time would be needed. Therefore, the shortcut described in section \ref{sec:shortcut} was used, where the central part of the PSF explicitly used in the inversions is indicated \lf{by a black square} in Figure~\ref{fig:hinodepsf}. Since the wings of the PSF are taken into account explicitly when calculating the defect and the merit function, they are properly included in the converged result.

Since the results did not really require it, no straylight contamination was included in the inversion, which is not to say that there is none in the data. However, since the fit to the Stokes profile with the lowest intensity in the inverted data cube, which was found to be around $0.12 I_{c,HSRA}$ in the darkest part of the umbra, was still accurate to within $\sim 0.1 I_{c,local}$, the amount of straylight present that cannot be explained with the telescope PSF is unlikely to be more than around $0.01 I_{c,HSRA}$, although no effort was made to more precisely determine this number. 

A scan of AR10933, recorded on \lf{5 January 2007}, 11:54-12:27 UT was selected, showing a sunspot umbra, penumbral filaments\lf{,} and some solar granulation, very close to the center of the solar \lf{disk}. The FOV was selected since it contains a wide variety of atmospheric conditions and fine structure that are likely to benefit from a spatially coupled inversion. As in the simulations from section \ref{sec:simulations}, three nodes were placed at $^{10}\log{\tau}={-2.5,-0.9,0.0}$ to represent the solar atmosphere.

\subsection{Inverted observations}
Before the inversion results are discussed, it should be stated that the focus of this paper is a discussion of the coupled inversion method and not a discussion of the atmospheric structure that is the actual result of the inversion. Since discussing the latter would draw our attention away from the method, we 
\begin{figure}[htb]
\includegraphics[height=9.00cm,bb=10 0 254 250]{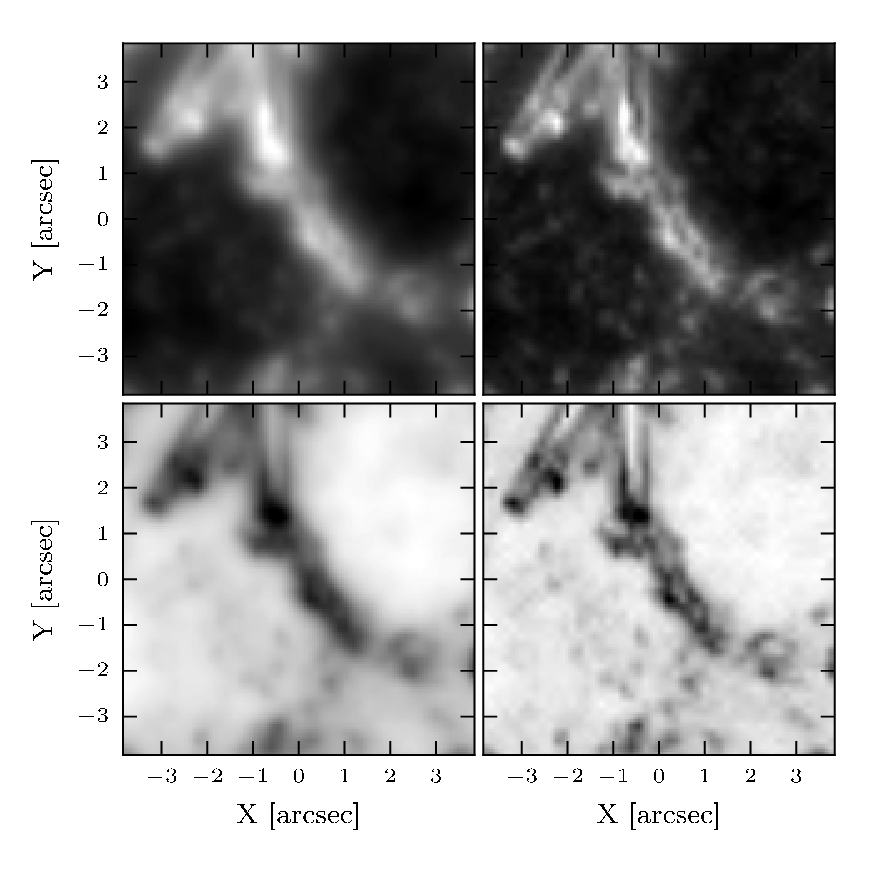}
\caption{Observed (left) and inverted (right) Stokes images in Stokes I (top) at and Stokes V (bottom) at 6302.55\AA.}
\label{fig:blowup1}
\end{figure}
leave a detailed analysis of \lf{these} results \lf{to} a separate paper and instead limit ourselves to a brief, qualitative assessment of the results combined with a comparison of the fitted undegraded profiles to the observed (degraded) data. More specifically, we produce images generated from the inverted profiles (inverted images) and compare them to images made from the observed data. 

The inverted images at 6302.55\AA\ of the selected active region are shown in Figure~\ref{fig:hinode2d}. The synthetic images generated from the inverted atmospheric structure (right side) compare very favorably with \begin{figure}[htb]
\includegraphics[height=8.00cm,bb=0 0 253 220]{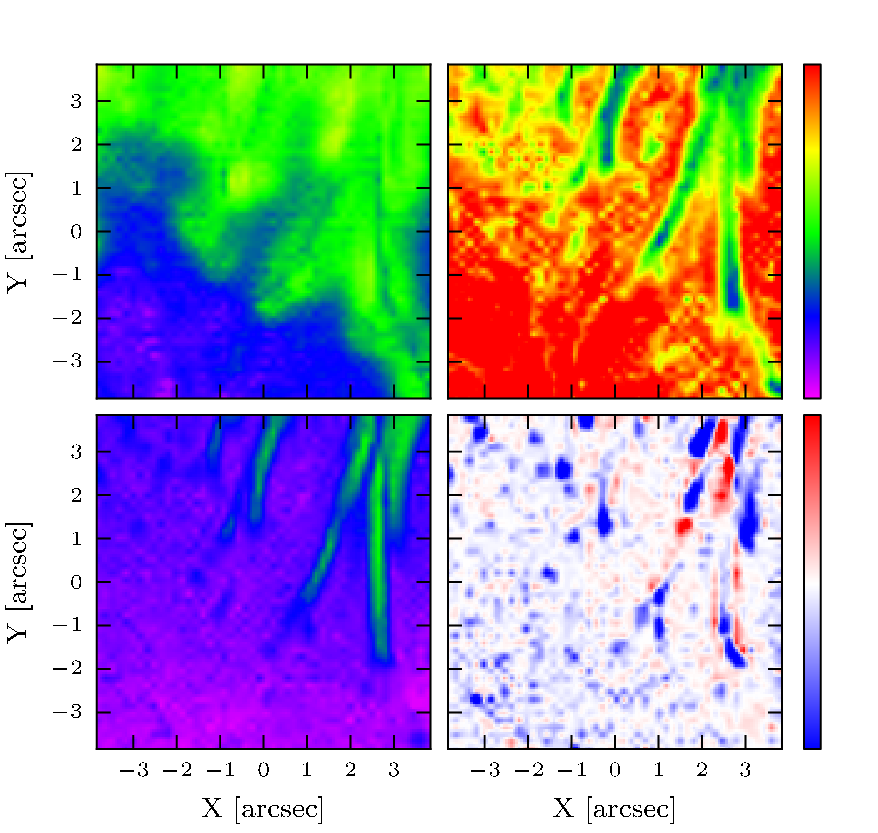}
\caption{A selection of inverted atmospheric parameters. From left to right, top to bottom: $T_{-2.5}$, $B_{-0.9}$, $\gamma_{-0.9}$, and $V_{D,0}$, with the subscript denoting the value of $^{10}{\log}\tau$. The color scale used for $T_{-2.5}$, $B_{-0.9}$, \lf{and} $\gamma_{-0.9}$ \lf{is} indicated on the top right, \lf{the one} for $V_{D,0}$ on the bottom right. The ranges of the color scale are [3300-5200]$\,$K, [500-2500]$\,$G, [0-180]$\,^\circ$\lf{,} and [-3,3]$\,km\,s^{-1}$\lf{,} respectively.}
\label{fig:blowup3}
\end{figure}
the original data (left side), in that they show much more contrast and a significant increase in the level of detail. This is even \lf{clearer} in Figure~\ref{fig:blowup1}, where the upper \lf{righthand} corner of the umbra is shown enlarged for both the observed and the inverted data.

A selection of atmospheric parameters is shown in Figure~\ref{fig:blowup3}, where the level of detail suffices to see the temperature drop in the penumbral dark cores at the highest node, the clear drop in magnetic field strength in the penumbral filaments, the large increase in inclination in the penumbral filaments\lf{,} and clear signs of \lf{downflow} on both sides of the filament in the deepest layers, \lf{only seen before} in datasets taken with larger solar telescopes \citep{2011ApJ...734L..18J,2011Sci...333..316S}.

Despite this obvious improvement in the inverted parameters, the results also show several regions with obvious signs of oscillations \lf{on} the scale of a single pixel. The oscillations \lf{show up} in specific regions and cannot be removed by offering the code a smooth start condition. The pattern itself is confined to a certain region and does not spread beyond that, but its phase is not fixed. Several inversions starting from slightly different start conditions all produce a similar oscillatory pattern in the same region, but none of them is identical to any of the others. An average of several such solutions does not show any oscillations, but if an inversion is started from this \begin{figure}[htb]
\includegraphics[height=9.00cm,bb=10 0 254 250]{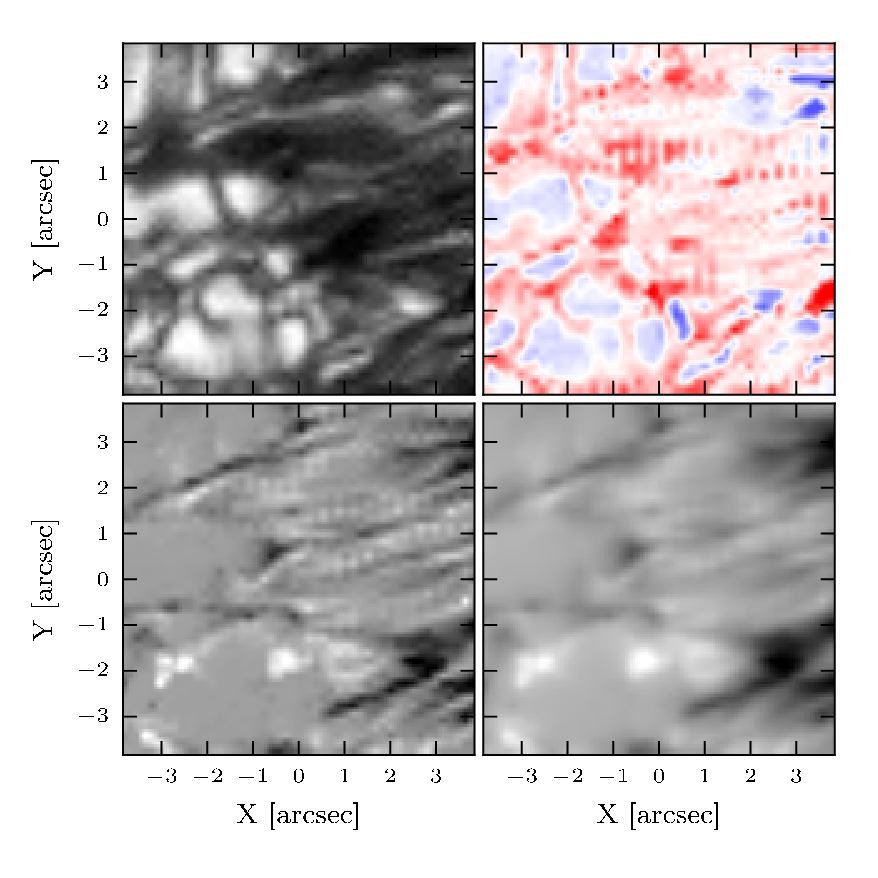}
\caption{Inverted Stokes images in Stokes I and V (left), alongside with the inverted LOS velocity at $\tau=1$ and the degraded Stokes V. Clearly visible are the oscillations in the penumbra, \lf{which} are invisible after degrading the image.}
\label{fig:blowup2}
\end{figure}
averaged solution, it will develop new oscillations.

The regions in which these oscillations are found are typically fine structured and fast evolving, such as the outer penumbra. One example is shown in Figure~\ref{fig:blowup2}, where the oscillations are clearly visible in the undegraded Stokes V and the LOS velocity in the penumbra, but are perfectly smooth in the granulation right next to it. The degraded Stokes V image shows that these oscillations average out to form a smooth image with only very weak oscillations.

The persistence of the oscillations in combination with the relative insensitivity of the inversion to their phase, but not their amplitude, suggest that we may not be dealing with just any degenerate solution here, but that the oscillations may actually be required to produce a spectral feature that cannot be produced by the atmospheric model. In particular, since the occurrence appears to be concentrated in the deep layers, the code must have \lf{difficulty} accurately reproducing the wings of the line profiles. This may be due to calibration errors, but it is more likely caused by \lf{fine-scale} structures that are not resolved by the telescope, possibly in combination with significant time evolution effects occurring during the spatial scan.

Apparently the inversion algorithm prefers to compensate for \lf{these} errors by introducing what is in effect a \lf{microturbulent} velocity structure, that is, by alternating the LOS velocity on the smallest possible scale. The averaging occurring in the convolution of the synthesized profiles then produces the broadened profiles that are observed. Interestingly, the removal of the micro-turbulent velocity as a fit parameter significantly increases the problem, suggesting that this parameter is used to fit some of the broadening caused by unresolved structure, but that significant differences remain that cannot be reproduced in this way.

The temperature map in Figure~\ref{fig:blowup3} shows some horizontal stripes that can clearly not be associated with real solar features. They are parallel to the spectrograph scanning direction and can only be attributed to imperfections in the \lf{flatfields}, made visible here \lf{by} the level of detail to which the observed data are fitted. These artifacts are concentrated in the top layer of the inverted atmosphere, most likely \lf{owing} to the small number of spectral points that are available to constrain this part of the atmosphere and the intrinsically lower intensity that these points have compared to the other parts of the spectrum.

Even a small calibration error in the width of the instrumental transmission profile will cause compensation effects in the solution, predominantly in the top layer. Indeed, a deliberate underestimate of the spectral transmission profile of the instrument introduces significant oscillations in the top layer of the atmosphere, as the algorithm attempts to broaden the line by alternating an atmospheric quantity.

It is clear that the algorithm is able to successfully fit the observed spectra, but it is able to do so with such high accuracy that it is essential to calibrate the data to a precision level that matches or exceeds that of the inversion result, which is significantly higher than the noise level of the data.

\subsection{Deconvolved observations}
\begin{figure}[htb]
\includegraphics[height=9.00cm]{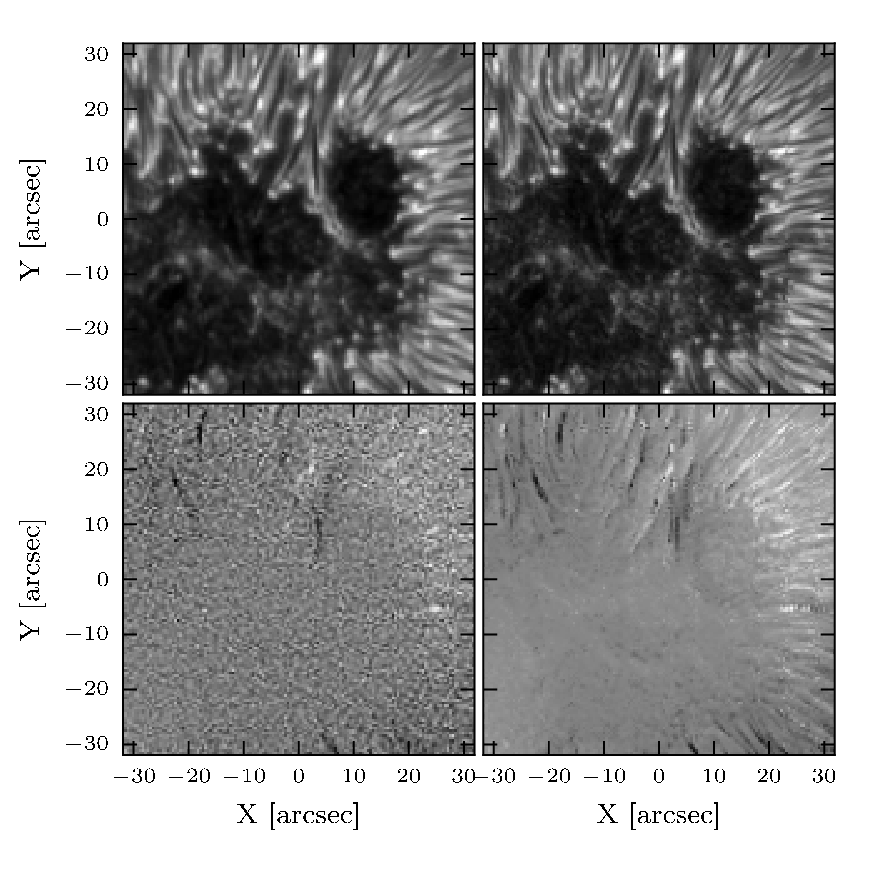}
\caption{Deconvolved (left) and inverted (right) Stokes images in Stokes I (top) at 6302.55\AA\ (line wing) and Stokes U (bottom) at 6302.65\AA\ (near continuum). Note the enhanced contrast in the high signal case (top) and the improved noise characteristics in the low signal case (bottom) of the inverted result as compared to the deconvolved result.}
\label{fig:deconvinvcomp}
\end{figure}
\begin{figure}[htb]
\includegraphics[height=3.00cm,bb=0 0 253 90]{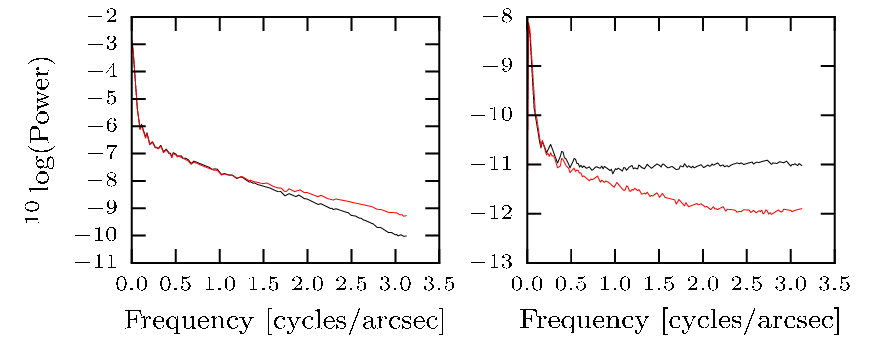}
\caption{Angular averaged power spectrum of the deconvolved (black) and inverted (red) solutions of Stokes I at 6302.55\AA, approximately half way into the Fe I line at 6302.5\AA. Clearly visible is the increased power of the inverted result in the \lf{high-frequency} range of the power spectrum.}
\label{fig:deconvinvpwr}
\end{figure}
To evaluate the results obtained with the coupled inversions, the inverted undegraded profiles were compared to the data deconvolved using Richardson-Lucy deconvolution \citep{1972JOSA...62...55R,1974AJ.....79..745L}. Figure~\ref{fig:deconvinvcomp} shows two images, one deconvolved and the other one the result of the coupled inversion. The Stokes I images, at  6302.55\AA, show a significant difference in contrast, as is confirmed by the azimuthally averaged power spectrum shown in Figure~\ref{fig:deconvinvpwr}, indicating that the inverted image contains much more power at the high frequencies than the deconvolved result. 

A \lf{low-signal} image, on the other hand, such as the Stokes U image shown at the bottom of Figure~\ref{fig:deconvinvcomp}, shows a power spectrum that has much more power than the inverted result, as can be seen in the \lf{righthand} panel of Figure~\ref{fig:deconvinvpwr}.
Close inspection of the deconvolved image confirms that this can be attributed to a high level of noise in the image, not to any actual structure. The inverted result\lf{,} on the other hand\lf{,} still shows credible, coherent structures \lf{with nine orders of magnitude less power than the maximum power in the Stokes I image} and does not contain any noise.

The inversion is able to produce \lf{high-frequency} information when it is supported by the data, but suppresses it very effectively elsewhere. The likely reason for this lies in the fact that the inversion is based on the complete data cube, whereas the deconvolved result \lf{only} contains information from a single wavelength. The effective noise level of the deconvolved result is therefore much higher than for the inverted result, specifically at very low signal levels, where the inverted result is constrained by the high-signal regions of the spectra, where noise is much less important.

\section{Conclusions}
We have developed a new method for \lf{inverting 2-D} maps of spectro-polarimetric data that have been degraded spatially in a known way. The method uses the information contained in the spectral dimension and the known spatial degradation properties to constrain a parameterization of the atmosphere over the whole FOV simultaneously. 

The method was formulated in a manner specific to an inversion code using a ``greedy'' optimization method\lf{,} and although it is implemented for the SPINOR inversion code, it should not be considered \lf{as} specific to this code. The method only changes the way in which the merit function is evaluated and modifies the response functions accordingly, implying that it should be easy to adapt to other inversion codes. The suitability of the specific implementation described in this paper is limited to the specific category of minimum search algorithms and cannot be \lf{used directly} in Monte-Carlo or genetic codes without adaptations.

Tests carried out on simulated MHD cubes show that on a critically sampled grid, the code is able to retrieve atmospheric parameters by inverting profiles synthesized from them and degraded using a realistic PSF, with only a small increase \lf{in} the error \lf{of} the retrieved parameters as compared to a pixel-by-pixel inversion of the undegraded spectra. For oversampled data, the code is able to successfully recover the \lf{low-frequency} components of the solution, but appears to have \lf{difficulty} recovering information beyond the diffraction limit\lf{. A}t no time \lf{was} unstable behavior resulting from oversampling the solution observed.

The method developed here is a specific case of a more generalized approach to data reduction, involving the application of calibration information to synthetic data (e.g. in the forward direction), instead of ``correcting'' the real data with the inverse calibration. \lf{Thanks} to this forward application of the calibration information, the noise amplification typically associated with the correction of degraded data is strongly reduced, since the inverted calibration data are never applied to it. In addition, by combining all the available information to constrain a simplified parameterized atmosphere, all errors and limitations are accumulated in one place, resulting in a robust solution with considerably improved specifications over the input data.


Based on the results of section \ref{sec:hinode} we conclude that this method is suitable for application to spectral imaging data, where the assumption of a constant PSF can be made. There is no fundamental problem extending \lf{this} method to the treatment of data where this assumption is not valid, which will be explored in a future paper.


\acknowledgements{The author wants to thank A. Lagg and S. Tiwari for the many helpful discussions that have resulted in the work presented here.}

\bibliography{ms}

\end{document}